\documentclass[twocolumn,showpacs,preprintnumbers,amsmath,amssymb]{revtex4}
\usepackage{hyperref}
\usepackage{floatrow}
\usepackage{graphicx}
\usepackage{dcolumn}
\usepackage{amsmath}
\usepackage{epsfig}
\usepackage{subfig}
\usepackage{array}
\usepackage{xcolor}
\usepackage{bm}

\begin{document}

\normalsize

\title{
The remaining parts for the long-standing $J/\psi$ polarization puzzle
}

\author{Yu Feng$^{1,2}$}
\author{Bin Gong$^{3,4}$}
\author{Chao-Hsi Chang$^{2,4}$}
\author{Jian-Xiong Wang$^{3,4}$}
\affiliation{
$^{1}$ Department of Physics, College of Basic Medical Sciences, Army Medical University, Chongqing 400038, China \\
$^{2}$ CAS Key Laboratory of Theoretical Physics, Institute of Theoretical Physics, Chinese Academy of Sciences, Beijing 100190, China\\
$^{3}$ Institute of High Energy Physics, Chinese Academy of Sciences, P.O.Box 918(4), Beijing 100049, China \\
$^{4}$School of Physical Sciences, University of Chinese Academy of Sciences, Beijing 100190, China
}

\begin{abstract}
Based on the non-relativistic quantum chromodynamics factorization formalism, 
the polarization parameters $\lambda_{\theta\phi}$ and $\lambda_{\phi}$ of $J/\psi$
hadroproduction are analyzed in helicity frame and
calculated at QCD next-to-leading order for the first time.
For prompt $J/\psi$ production, we take into account the feeddown contributions from $\chi_{cJ}$
and $\psi(2S)$ decays. The theoretical predictions for the polarization parameters $\lambda_{\theta\phi}$ and
$\lambda_{\phi}$ of $J/\psi$ are presented.
With the theoretical results we have done
the fit to the experimental measurements on yield and polarization for $J/\psi$
hadroproduction simultaneously, and found that the results are coincide with
the experimental measurements at the LHC quite well. 
\end{abstract}


\maketitle

{\it \textbf{Introduction}---}
Nonrelativistic quantum chromodynamics (NRQCD)~\cite{Bodwin:1994jh} is one of the most successful effective theory
to describe the decay and production of heavy quarkonium (as a review, see e.g.~\cite{Brambilla:2010cs,Brambilla:2014jmp}).
By separating the processes related to heavy quarkonium as short-distance coefficients (SDC)
and the long-distance matrix elements (LDMEs),
NRQCD allows one to organize perturbative calculations as double expansions in both the coupling constant $\alpha_s$
and the heavy quark relative velocity $v$.
In the past decade, great improvements are made in the next-to-leading order (NLO) QCD correction calculation
~\cite{Campbell:2007ws,Gong:2008sn,Gong:2008ft,Butenschoen:2010rq,Ma:2010yw,Butenschoen:2011yh,Ma:2010jj}.
The NLO corrections to color-singlet $J/\psi$ hadroproduction have been investigated in Ref.~\cite{Campbell:2007ws,Gong:2008sn},
and its transverse momentum ($p_t$) distribution is found to be enhanced by 2-3 orders of magnitude at high $p_t$ region
and the $J/\psi$ polarization changes from transverse into longitudinal at NLO~\cite{Gong:2008sn}.
The NLO corrections to $J/\psi$ production via S-wave color-octet (CO) states are first studied in Ref.~\cite{Gong:2008ft}
and the corrections to $p_t$ distributions of $J/\psi$ yield and polarization are small.
Ref.~\cite{Butenschoen:2010rq,Ma:2010yw,Butenschoen:2011yh,Ma:2010jj} performed
the complete NLO calculation for prompt $J/\psi$ hadroproduction and their results can fit the $p_t$ distributions
of the experimental measurements at Tevatron and LHC.

Despite all the successes, we cannot overlook the challenges it is facing.
In the $J/\psi$ case, the determination of CO long-distance matrix elements (LDMEs)
suffers ambiguity by the freedom of fitting method.
With different fitting strategies, quite different values of LDMEs are obtained, which lead to different descriptions
of the polarization distribution.
Three groups~\cite{Butenschoen:2012px,Chao:2012iv,Gong:2012ug} have made great efforts to proceed the calculation of $J/\psi$ polarization
$\lambda_\theta$ to QCD NLO,
but none of their CO LDMEs can reproduce the experimental measurements form LHC~\cite{Aaij:2013nlm,Aaij:2014qea} with good precision
for low and high $p_t$ range of $J/\psi$ simultaneously. The big uncertainty on the related LDMEs
still remains, and it is even more complicated that only one~\cite{Gong:2012ug} of the three groups
includes the $\chi_c$ and $\psi(2s)$ feeddown from direct calculation and the result can be compared with
the experimental measurement on prompt $J/\psi$ polarization.  Thereafter,
the $\eta_c$ hadroproduction measured by LHCb Collaboration~\cite{Aaij:2014bga} provides another laboratory
for the study of NRQCD. Ref.~\cite{Butenschoen:2014dra} considers it as a challenge to NRQCD,
while Refs.~\cite{Han:2014jya, Zhang:2014ybe} found these data are consistent with the $J/\psi$ hadroproduction measurements.
Further, with the constraint on the LDMEs obtained in Ref.~\cite{Zhang:2014ybe,Sun:2015pia}, the authors
found a special way to reduce the LDMEs uncertainty for the $J/\psi$.

The $J/\psi$ polarization, encoded in the angular distributions of the lepton pair,
is described by
\begin{equation}\label{equ:decay}
\begin{split}
\frac{d^2N}{d \cos\theta d\phi}\propto
1&+\lambda_{\theta}\cos^2\theta + \lambda_{\theta\phi}\sin(2\theta)\cos\phi \\
&+\lambda_{\phi}\sin^2\theta\cos(2\phi)
\end{split}
\end{equation}
where $\theta$ is the polar angle between the direction of the positive lepton and chosen polarization axis,
and $\phi$ is the azimuthal angle, measured with respect to the production plane.
While all the three coefficients provide independent information, almost all the theoretical studies of $J/\psi$ polarization
are restricted to $\lambda_{\theta}$.
The parameter $\lambda_{\phi}$  has only been studied at QCD NLO work in Ref.~\cite{Butenschoen:2012px}
with a few experimental data points measured by ALICE Collaboration~\cite{Abelev:2011md}.
For $\lambda_{\theta\phi}$, there is no theoretical prediction at all.
On the other side, there are experimental measurements on $\lambda_{\theta\phi}$ and $\lambda_{\phi}$  for
$J/\psi$ and $\psi(2S)$ polarization from the LHCb~\cite{Aaij:2013nlm,Aaij:2014qea} and the CMS~\cite{Chatrchyan:2013cla}.
Based on the results from all the theoretical studies, it is believed
that a combined fit includes $p_t$ distribution on $J/\psi$ production rate and polarization parameter $\lambda_\theta$ can
 be achieved. But it does not mean that the $J/\psi$ polarization puzzle is solved. There are still two parameters $\lambda_{\theta\phi}$
and $\lambda_{\phi}$ from experimental measurements without theoretical predictions to compare with.
Therefore, theoretical study on these two polarization parameters is certainly needed. Are the theoretical predictions on these two parameters
coincide with the experimental measurements? Otherwise, could the uncertainty on the related LDMEs be reduced by fitting on these measurements
together with previous data fit? These are very important issue to settle down for the long-standing $J/\psi$ polarization puzzle.

In this Letter, we perform a theoretical analysis on the property of polarization parameters, and finish the
calculation on $\lambda_{\theta\phi}$ and $\lambda_{\phi}$
for $J/\psi$ and $\psi(2S)$ polarization in helicity frame based on NRQCD at QCD NLO.
The results are obtained for the first time. By performing a combined fit, they coincide with the experimental measurements at the LHC quite well.
Therefore, the last two pieces for the $J/\psi$ polarization are successfully explained.
It means that the long-standing $J/\psi$ polarization puzzle is settled down completely.

{\it \textbf{Calculation}---}
The three polarization parameters $\lambda_{\theta}$, $\lambda_{\theta\phi}$ and $\lambda_{\phi}$ in Eq.~\ref{equ:decay}
are are defined as ~\cite{Beneke:1998re}
\begin{equation*}
\lambda_{\theta}=\frac{d\sigma_{11}-d\sigma_{00}}{d\sigma_{11}+d\sigma_{00}},
\lambda_{\theta\phi} = \frac{\sqrt{2}Re d\sigma_{10}}{d\sigma_{11}+d\sigma_{00}},
\lambda_{\phi} = \frac{d\sigma_{1,-1}}{d\sigma_{11}+d\sigma_{00}},
\end{equation*}
Here, $d\sigma_{\lambda \lambda'}$($\lambda, \lambda'$= $0,\pm1$) is the spin density matrix of $J/\psi$ ($\psi(2S)$) hadroproduction,
which following the NRQCD factorization, can be expressed as
\begin{equation}\label{eq:ppsig}
\begin{split}
d\sigma_{\lambda \lambda'}(pp\rightarrow &HX)=\sum_{a,b,n}\int dx_1 dx_2 f_{a/p}(x_1) f_{b/p}(x_2) \\
&\times d\hat{\sigma}_{\lambda \lambda'}(ab\rightarrow (c\overline{c})_nX)\langle{\cal O}^{H}_{n}\rangle
\end{split}
\end{equation}
where $p$ is proton, the index $a$, $b$ run over the gluon ($g$) and the light quarks ($q$) and anti-quarks ($\overline{q}$).
$n$ denotes the color, spin and angular momentum states of the $c\overline{c}$ intermediate states
($^3S_1^{[1]}$, $^1S_0^{[8]}$, $^3S_1^{[8]}$, $^3P_J^{[8]}$) for $J/\psi$ and $\psi(2S)$,
or ( $^3P_J^{[1]}$, $^3S_1^{[8]}$) for $\chi_{cJ}$.
The functions $f_{a/p}(x_1)$ and $f_{b/p}(x_2)$ are the parton distribution functions (PDFs) for the incoming protons for parton types $a$ and $b$.
The SDC $d\hat{\sigma}$ can be perturbatively calculated
and the LDMEs $\langle{\cal O}^{H}_{n}\rangle$ are governed
by nonperturbative QCD effects.

For inclusive $J/\psi$ production at the LHC,
P parity is invariance for initial states protons and it involve only QCD interaction
which is also P parity invariance. Based on P parity invariance
for the production density matrix of the $J/\psi$ hadroproduction,  together with definition of the helicity frame,
a symmetry (asymmetry) relations can be deduced as (the detail is presented in Ref~\cite{Fengyu:2018b}):
\begin{equation}\label{eq:asym}
\frac{d\sigma_{\lambda \lambda'}^{H}}{dy}\big|_{y=a}=n_{\lambda \lambda'}
\frac{d\sigma_{\lambda \lambda'}^{H}}{dy}\big|_{y=-a}, ~
n_{\lambda \lambda'}=\Big\{^{~1 ~~\lambda=\pm\lambda'}
_{-1 ~\lambda=\pm1, \lambda'=0}
\end{equation}
Then the conclusions are obtained as $\lambda_{\theta\phi}=0$ for experiment
with symmetry rapidity range ($a<|y|<b$) like that at the CMS and ATLAS,
$\lambda_{\theta\phi}\neq0$ for half rapidity range ($y>b$) such as the case at the LHCb,
and $\lambda_{\theta},\lambda_{\phi}$ is symmetric for positive and negative rapidity in helicity frame.

To include the feed-down contributions from $\psi(2S)$ and $\chi_{cJ}$ to $J/\psi$,
we following the same treatment as in Ref.~\cite{Gong:2012ug},
\begin{eqnarray}
&&d\sigma_{\lambda \lambda'}^{J/\psi}|_{\psi(2S)}={\cal B}[\psi(2S)\rightarrow J/\psi]d\sigma_{\lambda \lambda'}^{\psi(2S)}, \\
&&\begin{split}
d\sigma_{\lambda \lambda'}^{J/\psi}|_{\chi_{cJ}}=&{\cal B}[\chi_{cJ}\rightarrow J/\psi]\sum_{J_z,J'_z}
\delta_{J_z-\lambda,J'_z-\lambda'} \\
&\times C^{\lambda,J_z-\lambda}_{J,J_z}C^{*\lambda',J'_z-\lambda'}_{J,J'_z}d\sigma_{J_zJ'_z}^{\chi_{cJ}}.
\end{split}
\end{eqnarray}
where $C^{\lambda,J_z-\lambda}_{J,J_z}$ is the Clebsch-Gordan coefficient,
and $\lambda$, $J_z$ are the quantum numbers of angular momentum.

To calculate the NRQCD prediction on the transverse momentum $p_t$ distribution of yield and polarization
for heavy quarkonium hadroproduction at QCD NLO, we use the FDCHQHP package~\cite{Wan:2014vka},
which was based on the collection of the Fortran codes generated
for all 87 parton level sub-processes by using FDC package~\cite{Wang:2004du} and implementation tool on job submission
and numerical precision control. It is a very powerful tool for us to save a lot of work and finish this study. Even with
it, we found that there were a few places had been improved on job submission and numerical precision control,
and improved them. The updated version FDCHQHP~\cite{Fengyu:2018c} will be publicly available soon.

{\it \textbf{LDMEs Strategy}---}
The CS LDMEs are estimated from the wave functions at the origin by
\begin{eqnarray}
\nonumber\langle{\cal O}^{\psi}(^{3}S^{[1]}_{1})\rangle&=&\frac{3N_c}{2\pi}|R_{\psi}(0)|^{2}, \\
 \langle{\cal O}^{\chi_{cJ}}(^{3}P^{[1]}_{J})\rangle&=&\frac{3}{4\pi}(2J+1)|R'_{\chi_{c}}(0)|^{2}.
\end{eqnarray}
where the wave functions at the origin can be calculated via potential model~\cite{Eichten:1995ch},
which gives $|R_{J/\psi}(0)|^{2}$=0.81 GeV$^3$, $|R_{\psi(2S)}(0)|^{2}$=0.53 GeV$^3$, and $|R'_{\chi_c}(0)|^{2}$=0.075 GeV$^5$.

The CO LDMEs are extracted from the fit on experimental data with QCD NLO theoretical formula.
However, different results for the LDMEs are obtained when different strategy are used in the fit.
We briefly discuss different fit results and made a selection of them to represent
the uncertainty on predictions induced by the LDMEs in following.

In the $J/\psi$ case, several groups of LDMEs~\cite{Butenschoen:2011yh,Ma:2010yw,Chao:2012iv,
Gong:2012ug,Zhang:2014ybe,Han:2014jya,Bodwin:2014gia}  can be found.
They are extracted by fitting the data of hadroproduction yield~\cite{Ma:2010yw, Gong:2012ug}, or combined with
polarization~\cite{Chao:2012iv} on $pp$ collisions.
In their fits~\cite{Ma:2010yw,Chao:2012iv, Gong:2012ug}, the data with $p_t < 7$ GeV are excluded.
Ref.~\cite{Butenschoen:2011yh} extracted the LDMEs with a wider set of data including the lower $p_t$ region ($p_t>3$ GeV)
hadroproduction and the production at $ep$ and $\gamma\gamma$ colliders with $p_t > 1$ GeV.
By the assumption of heavy quark spin symmetry (HQSS),
the fit in Ref.~\cite{Zhang:2014ybe,Han:2014jya} took the $\eta_c$ measurement~\cite{Aaij:2014bga} ($p_t \ge 6$ GeV)
into consideration, and they obtained consistent $J/\psi$ LDMEs.
In Ref.~\cite{Bodwin:2014gia}, the authors incorporate the leading-power fragmentation corrections together with
the usual QCD NLO corrections, which involves different SDC and results in different LDMEs.
The group of Ref.~\cite{Ma:2010yw} improved their analysis by taking into account
the feed-down contributions later~\cite{Shao:2014yta}, but no updated $J/\psi$ LDMEs are presented.
Among the LDMEs set mentioned above, only Ref.~\cite{Gong:2012ug} fitted the prompt $J/\psi$ hadroproduction
by including the feed-down contributions from $\chi_{cJ}$ and $\psi(2S)$ which can be used to calculate these feed-down contribution
to $J/\psi$ polarization.

Following the same treatment in Ref.~\cite{Gong:2012ug},
we refit the $J/\psi$ LDMEs by using more experimental data where the data in $p_t<7$ GeV region are excluded as usual.
In addition to the $p_t$ distribution of
the production yield data from the CDF~\cite{Acosta:2004yw} and the LHCb~\cite{Aaij:2011jh} in old fit,
the polarization data of $\lambda_{\theta}$, $\lambda_{\theta\phi}$ and $\lambda_{\phi}$ from the LHCb~\cite{Aaij:2013nlm}
and $\lambda_{\theta}$ and $\lambda_{\phi}$ from the CMS~\cite{Chatrchyan:2013cla} are used. While
the $\lambda_{\theta\phi}$ is exactly zero in our calculation for the CMS so that it can not be fitted.
The non-zero data on $\lambda_{\theta\phi}$ for the CMS could be from P-parity broken, i.e. from the electro-weak
production for $J/\psi$. Therefore we calculated the leading contribution from $^3S_1^{[1]}$, $^1S_0^{[8]}$, $^3S_1^{[8]}$,
$^3P_J^{[8]}$ and found that their production rate is smaller about 5 order in magnitude than the QCD production processes.
So we suggest that $\lambda_{\theta\phi}$ for the CMS should be constrained as zero in the experimental measurement.

To deal with the feed-down contributions from $\psi(2S)$ and $\chi_{cJ}$,
we use the CO LDMEs from Ref.~\cite{Gong:2012ug}, namely
$\langle{\cal O}^{\psi(2S)}(^{1}S^{[8]}_{0})\rangle$=-1.2$\times$10$^{-4}$ GeV$^3$,
$\langle{\cal O}^{\psi(2S)}(^{3}S^{[8]}_{1})\rangle$=3.4$\times$10$^{-3}$ GeV$^3$,
$\langle{\cal O}^{\psi(2S)}(^{3}P^{[8]}_{0})\rangle/m_Q^2$ =4.2$\times$10$^{-3}$ GeV$^3$,
and $\langle{\cal O}^{\chi_{c0}}(^{3}S^{[8]}_{1})\rangle$= 2.21$\times$10$^{-3}$ GeV$^3$.

By fitting the totally 86 data points of $J/\psi$ and minimizing $\chi^2$, we obtain
\begin{equation}\label{eq:ldmes}
\begin{split}
&\langle{\cal O}^{J/\psi}(^{1}S^{[8]}_{0})\rangle=(5.66\pm0.47)\times10^{-2} GeV^3,\\
&\langle{\cal O}^{J/\psi}(^{3}S^{[8]}_{1})\rangle=(1.17\pm0.58)\times10^{-3} GeV^3,\\
&\langle{\cal O}^{J/\psi}(^{3}P^{[8]}_{0})\rangle/m_Q^2=(5.4\pm0.5)\times10^{-4} GeV^3,
\end{split}
\end{equation}
which will be taken as default values to present our results on polarization parameters.

To investigate the uncertainties from different set for the values of LDMEs,
the other five sets of LDMEs in Table.~\ref{tab:ldmes-jpsi} are also used to present the final
numerical results.

\begin{table*}
  \caption[]{The values of LDMEs for $J/\psi$(in units of GeV$^3$). }
  \begin{tabular*}{\columnwidth}{@{\extracolsep{\fill}}l*{4}{c}}
  \hline\hline
   Reference & $\langle{\cal O}^{J/\psi}(^{3}S^{[1]}_{1})\rangle$ ~&~ $\langle{\cal O}^{J/\psi}(^{1}S^{[8]}_{0})\rangle$
  ~&~ $\langle{\cal O}^{J/\psi}(^{3}S^{[8]}_{1})\rangle$ ~&~ $\langle{\cal O}^{J/\psi}(^{3}P^{[8]}_{0})\rangle/m_Q^2$  \\
  \hline
  Butenschoen \textit{et al}.(2011)~\cite{Butenschoen:2011yh} ~&~ 1.32 ~&~ $3.04\times 10^{-2}$ ~&~ $1.68\times 10^{-3}$ ~&~ $-4.04 \times 10^{-3}$  \\
  Chao \textit{et al}.(2012)~\cite{Chao:2012iv}   ~&~  1.16 ~&~ $8.9\times 10^{-2}$ ~&~ $ 3.0\times 10^{-3}$  ~&~ $ 5.6 \times 10^{-3}$  \\
  Gong \textit{et al}.(2013)~\cite{Gong:2012ug} ~&~  1.16 ~&~ $9.7\times 10^{-2}$ ~&~ $-4.6\times 10^{-3}$  ~&~ $-9.5\times 10^{-3}$  \\
  Bodwin \textit{et al}.(2014)~\cite{Bodwin:2014gia} ~&~ 0 ~&~ $9.9\times 10^{-2}$ ~&~ $1.1 \times 10^{-2}$ ~&~  $4.9\times 10^{-3}$ \\
  Zhang \textit{et al}.(2015)~\cite{Zhang:2014ybe} ~&~ $0.645$  ~&~ $0.78\times10^{-2}$ ~&~ $1.0\times 10^{-2}$ ~&~ $1.7\times10^{-2}$ \\
  \hline\hline
  \end{tabular*}\label{tab:ldmes-jpsi}
  \end{table*}

{\it \textbf{Numerical results}---}
In our numerical calculation, the parton distribution function CTEQ6M ~\cite{Pumplin:2002vw}
and the corresponding two-loop QCD coupling constant $\alpha_s$ are used.
The charm-quark mass is chosen as $m_c$=$M_H/2$ approximately,
where the masses of relevant quarkonia $M_H$ are 3.0GeV, 3.5GeV and 3.686GeV
for $H$=$J/\psi$, $\chi_{cJ}$($J$=0,1,2) and $\psi(2S)$, respectively.
The renormalization and factorization scales are chosen as
$\mu_r$=$\mu_f$=$\sqrt{4m_c^2+p_t^2}$, while the NRQCD scale is $\mu_{\Lambda}$=$m_c$.
Branching ratios are ${\cal B}[\psi(2S)\rightarrow J/\psi]$=0.61 and
${\cal B}[\chi_{cJ}\rightarrow J/\psi]$=0.0127, 0.339, 0.192
for $J=0,1,2$~\cite{Patrignani:2016xqp}, respectively.
Additionally, a shift $p_t^H\approx p_t^{H'}\times(M_H/M_{H'})$ is used while considering the
kinematics effect in the feeddown from higher excited states.

\begin{figure}[htb!]
\centering
  \includegraphics[width=0.45\textwidth,origin=c]{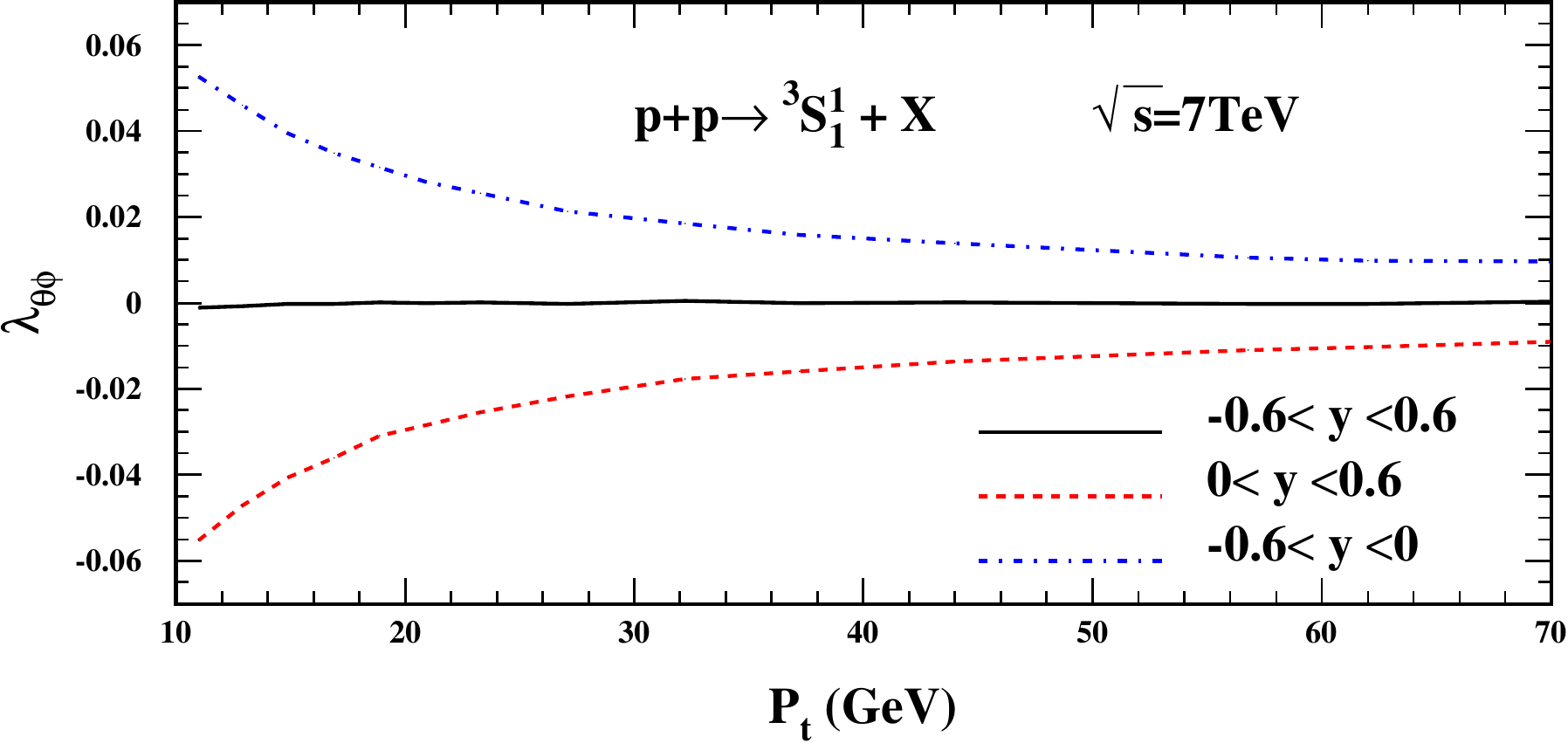}
  \includegraphics[width=0.45\textwidth,origin=c]{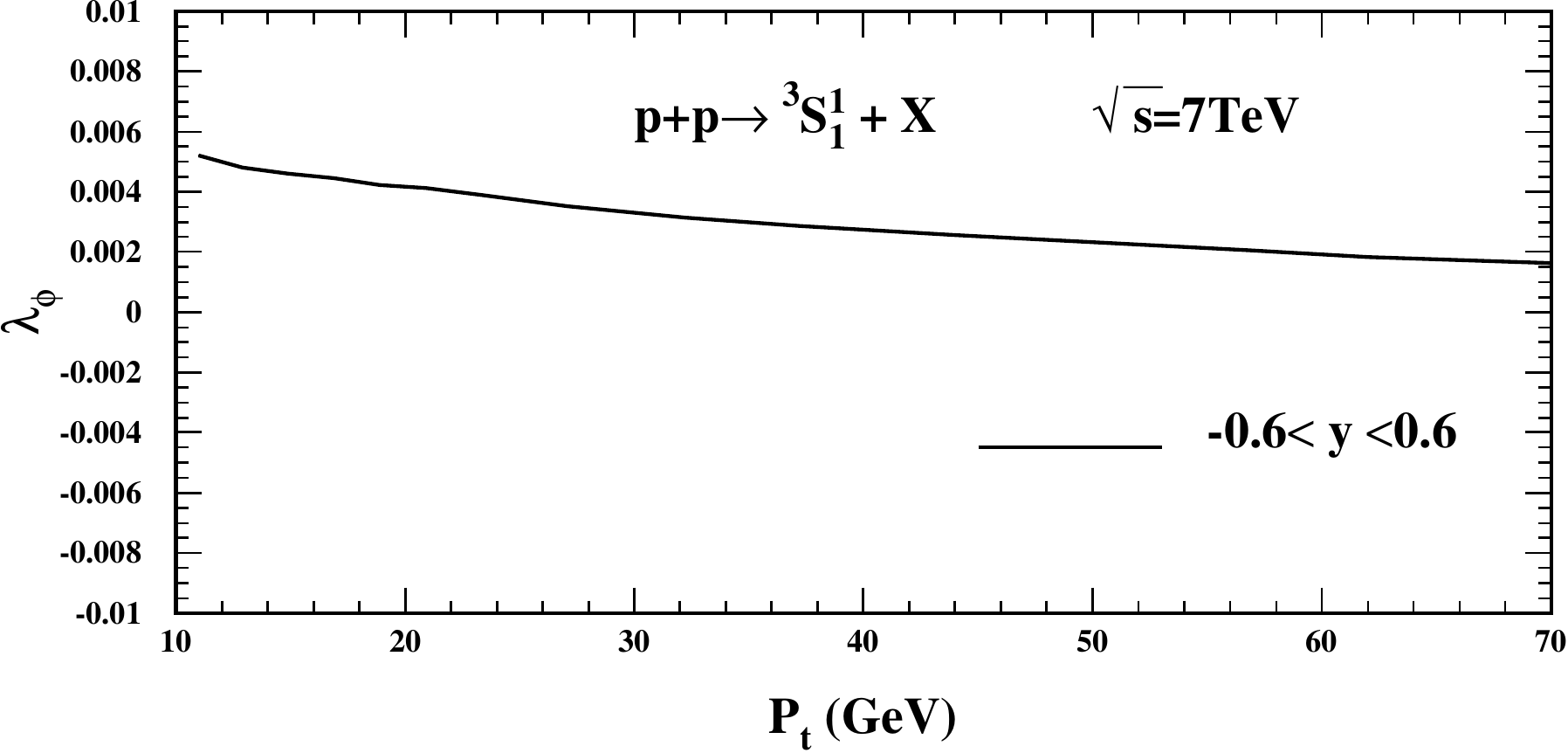} \\
  \includegraphics[width=0.45\textwidth,origin=c]{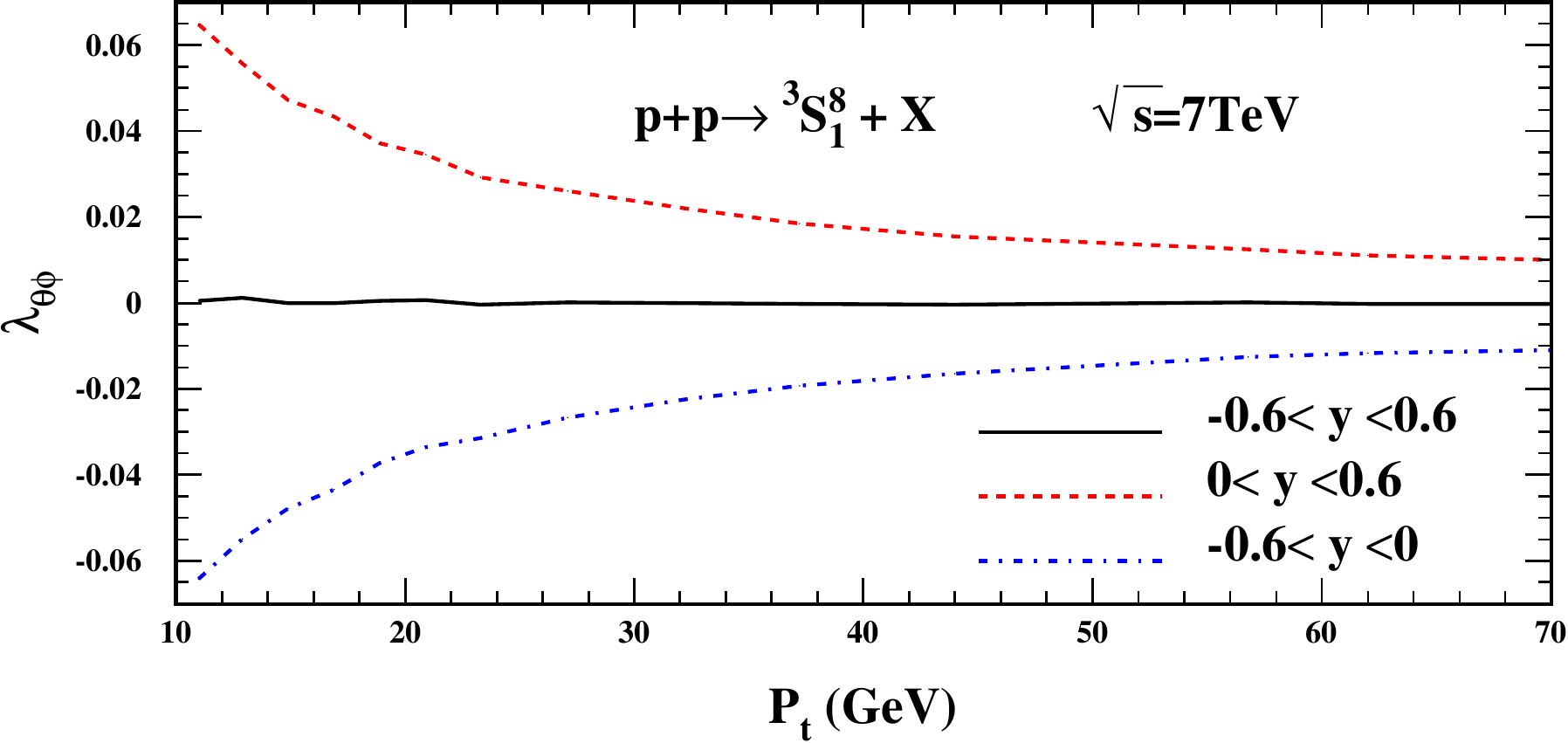}
  \includegraphics[width=0.45\textwidth,origin=c]{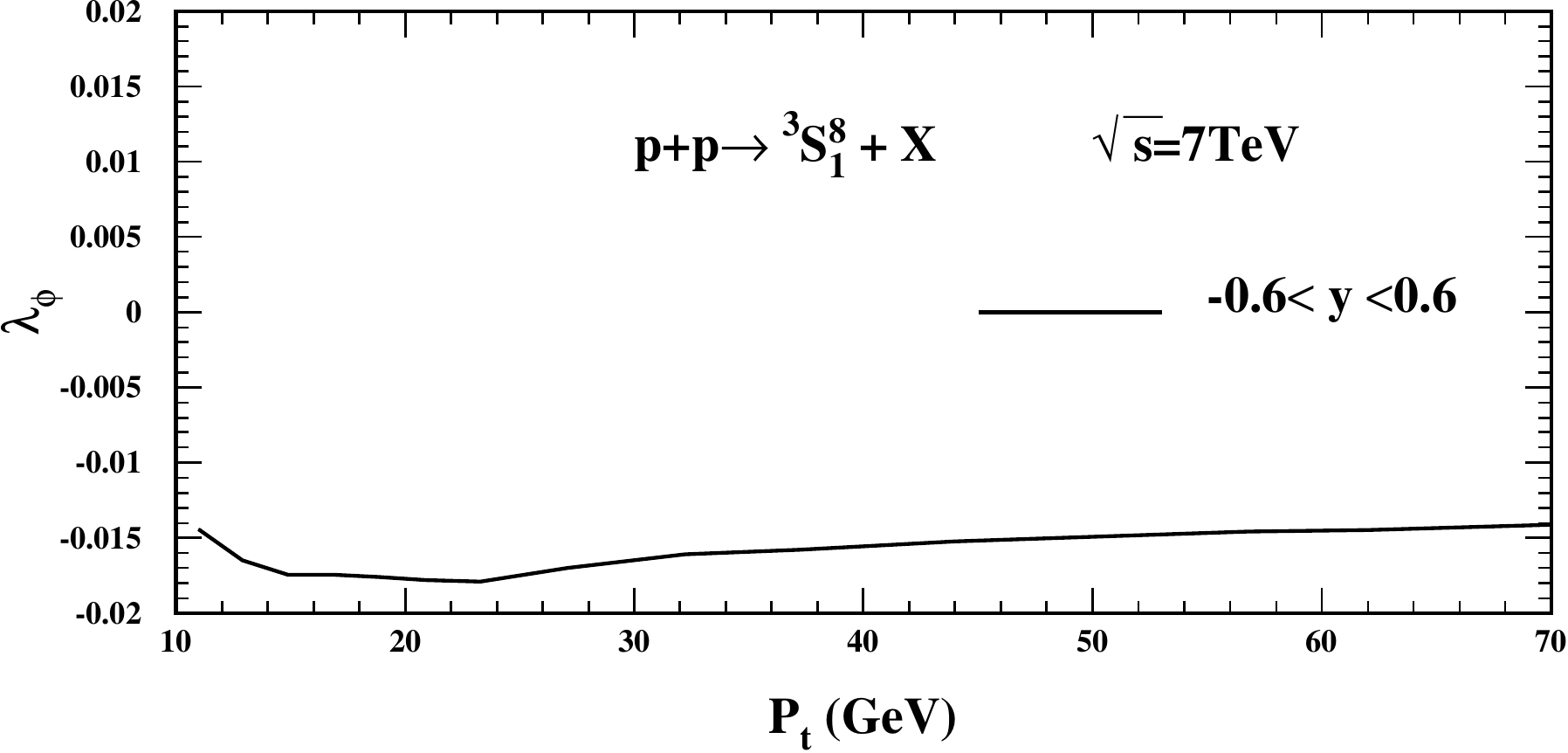} \\
  \includegraphics[width=0.45\textwidth,origin=c]{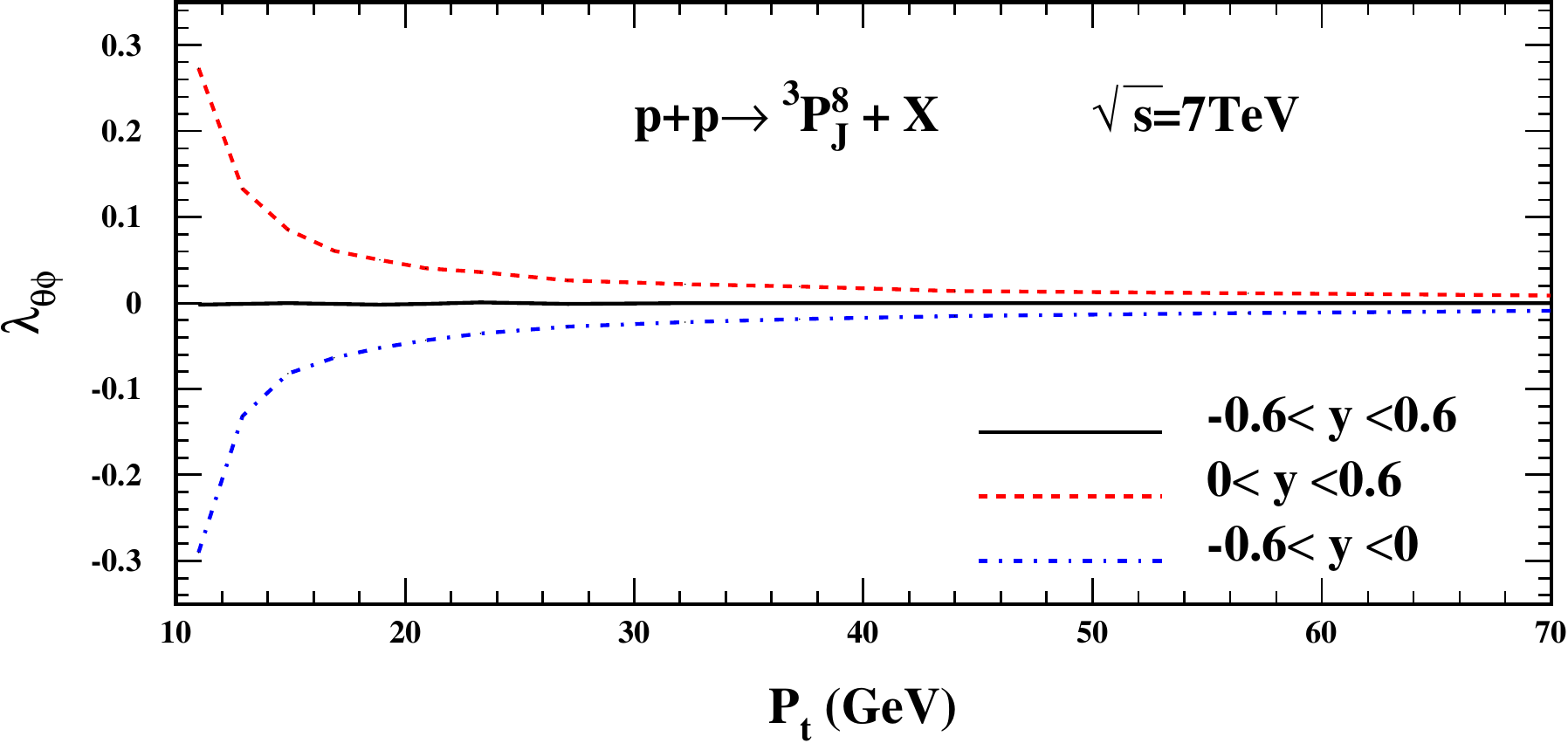}
  \includegraphics[width=0.45\textwidth,origin=c]{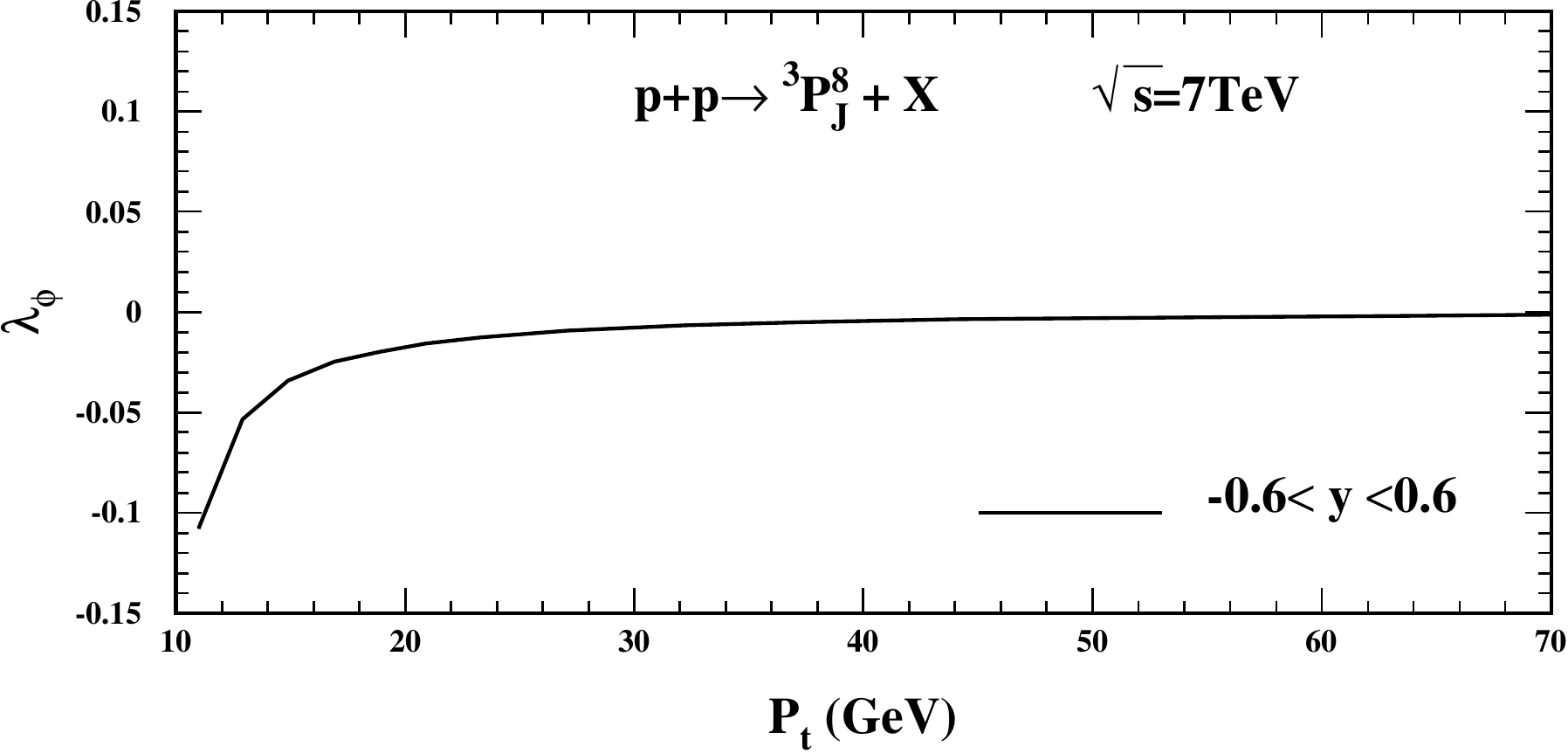}
  \caption{$\lambda_{\theta\phi}$ (left column) and $\lambda_{\phi}$ (right column)of $J/\psi$ with rapidity region being separated as the positive one (red dashed line) and the negative one (blue dash-dotted line).
  \label{fig:cms-chan} }
\end{figure}

\begin{figure}[htb!]
\centering
  \includegraphics[width=0.45\textwidth,origin=c]{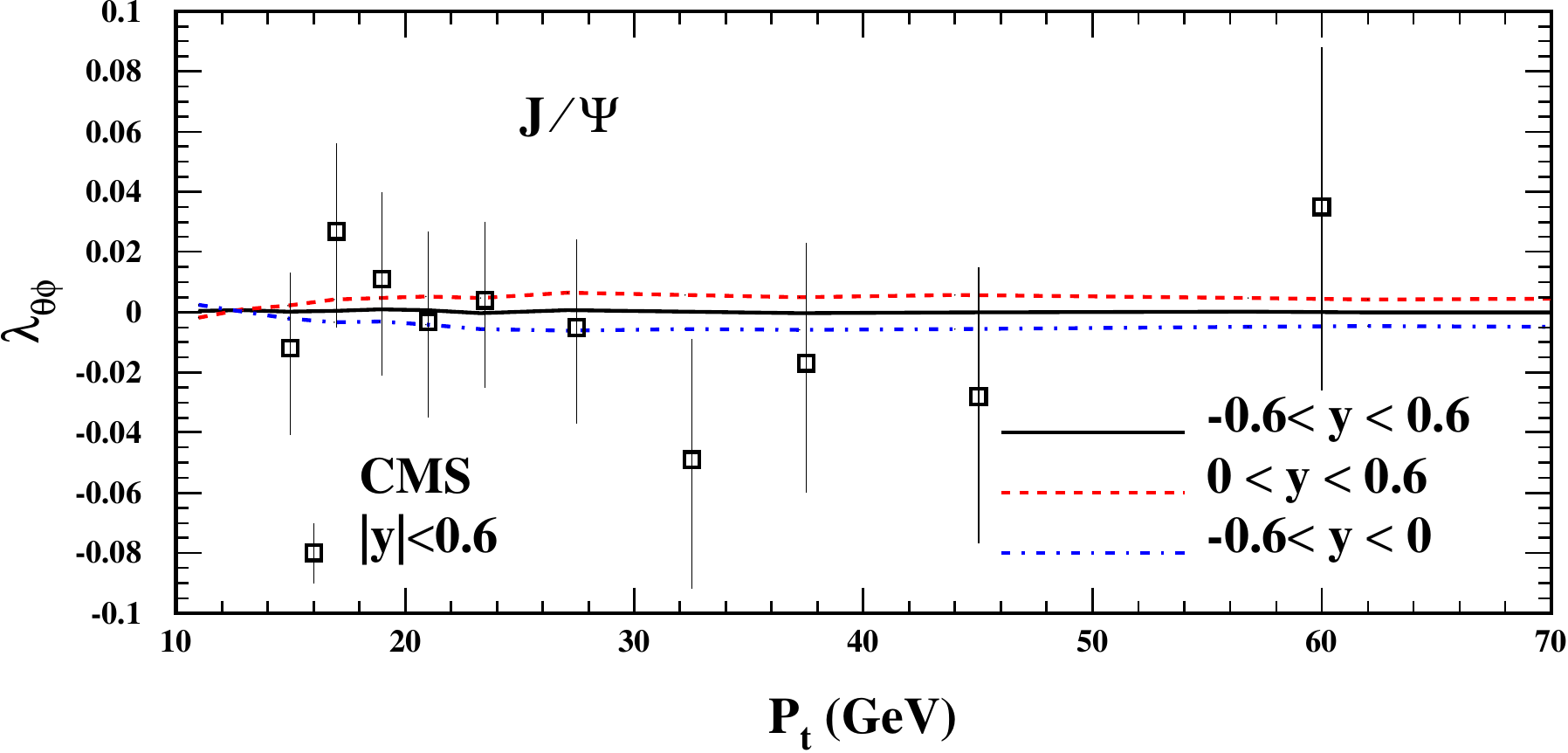}
  \includegraphics[width=0.45\textwidth,origin=c]{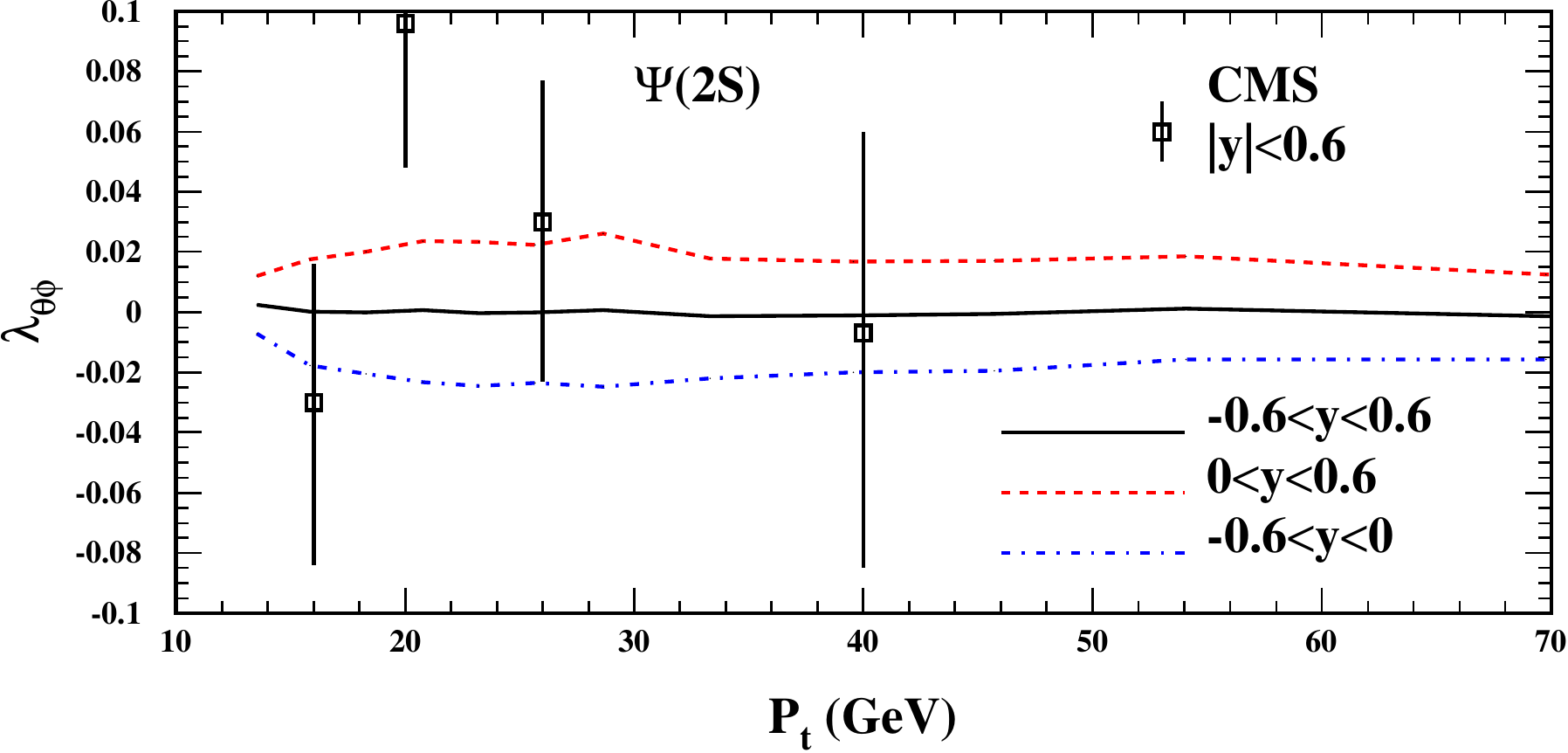}\\
  \includegraphics[width=0.45\textwidth,origin=c]{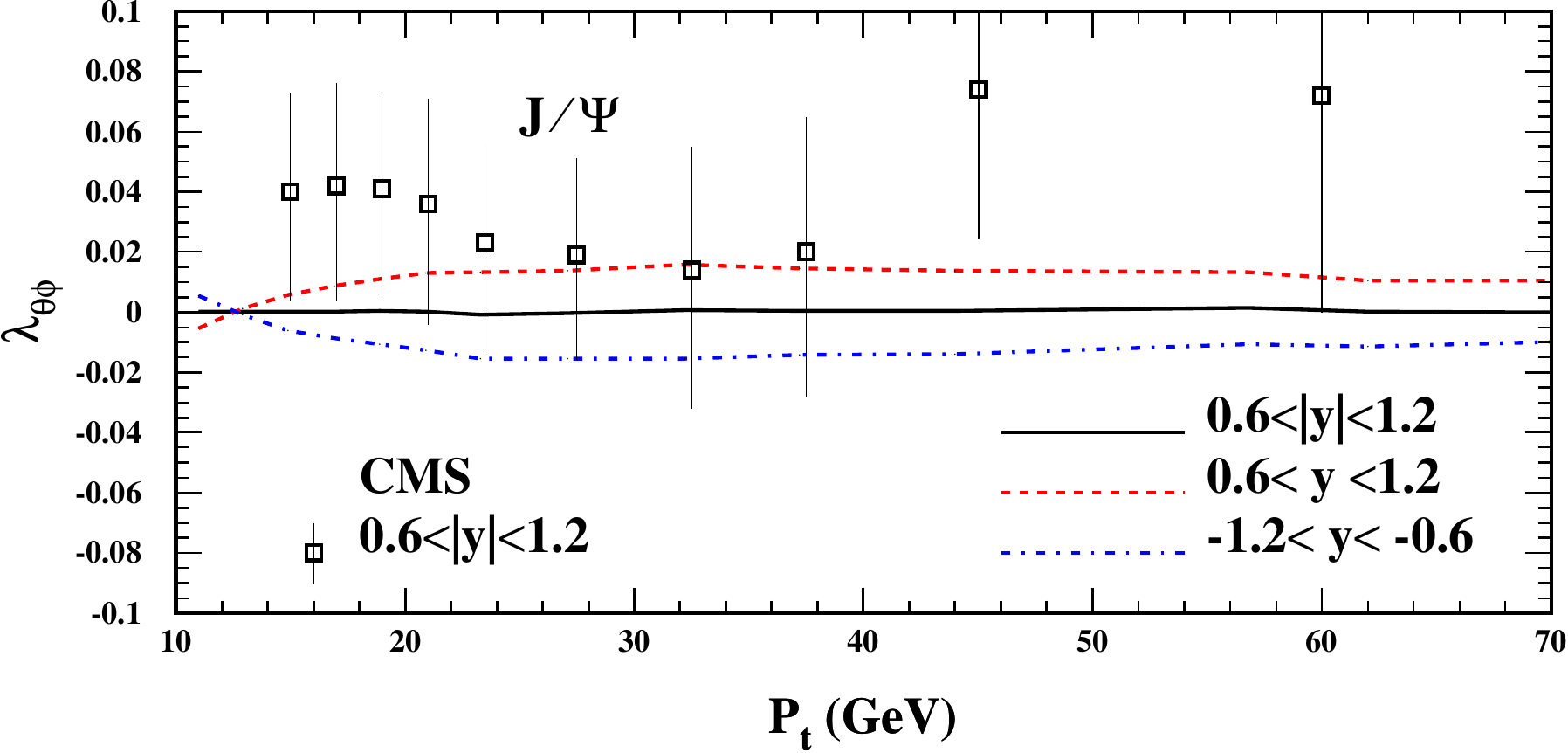}
  \includegraphics[width=0.45\textwidth,origin=c]{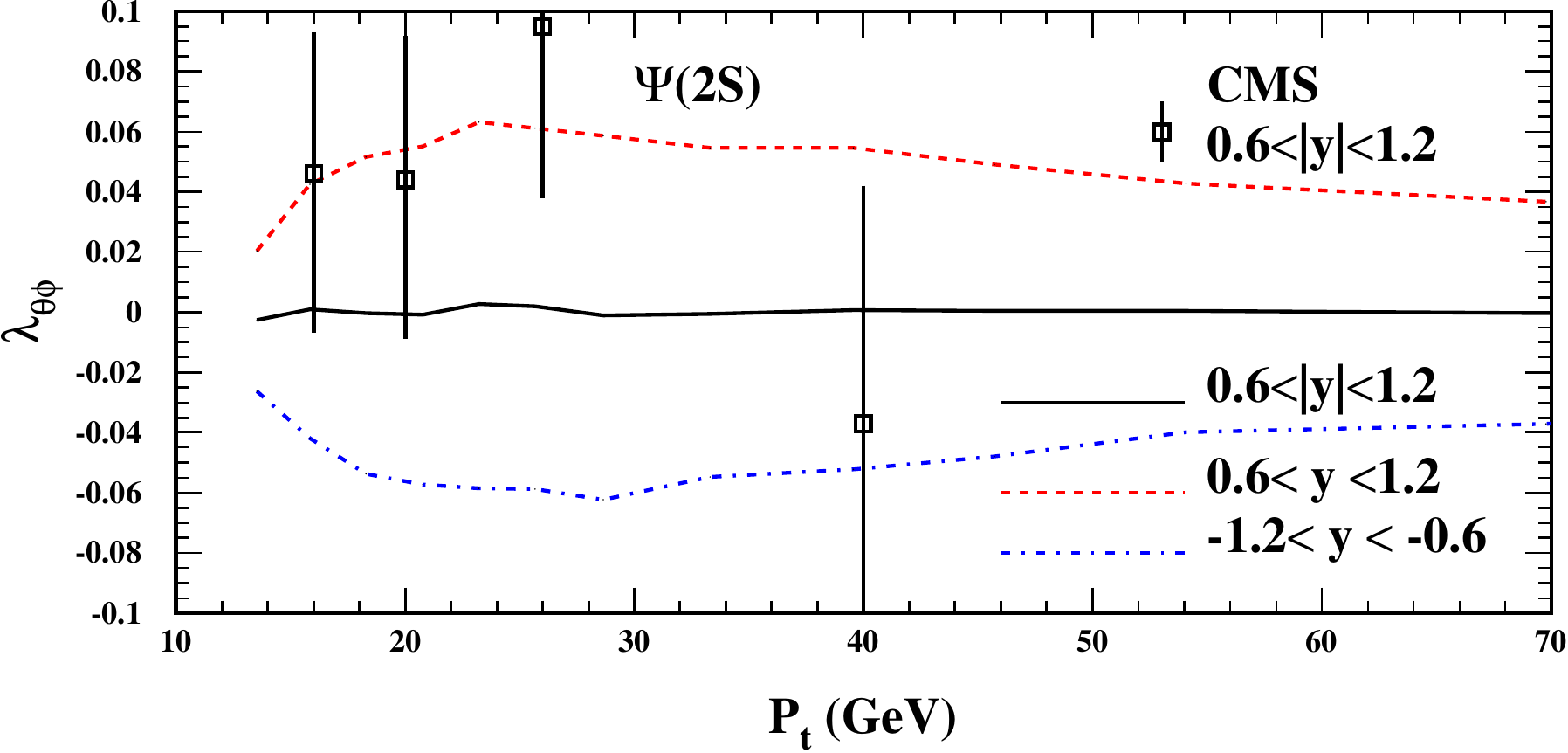}\\
  \caption{$\lambda_{\theta\phi}$ of $J/\psi$ (left column)
  and $\psi(2S)$ (right column) with rapidity region being separated as the positive one (red dashed line)
  and the negative one (blue dash-dotted line). The CMS data is from Ref.~\cite{Chatrchyan:2013cla}.
  \label{fig:cms-y} }
\end{figure}

In Fig.~\ref{fig:cms-chan}, $\lambda_{\theta\phi}$ is presented for each  ($^3S_1^{[1]}$, $^3S_1^{[8]}$, $^3P_J^{[8]}$) channel with rapidity range $y>0$, $y<0$ and $|y|>0$,
and it shown that $\lambda_{\theta\phi}$ is of asymmetry for rapidity range just as the conclusion from Eq.\ref{eq:asym}.

By using the default LDMEs set in Eq.~\ref{eq:ldmes}, $\lambda_{\theta\phi}$ at the CMS
is presented in Fig.~\ref{fig:cms-y}.
It shown that final numerical results are of asymmetry within very good numerical precision
and the theoretical prediction coincide with the experimental measurements at the CMS quite well
for both $J/\psi$ and $\psi(2S)$.

\begin{figure}[tbp]
  \centering
  \includegraphics[width=0.3\textwidth,origin=c]{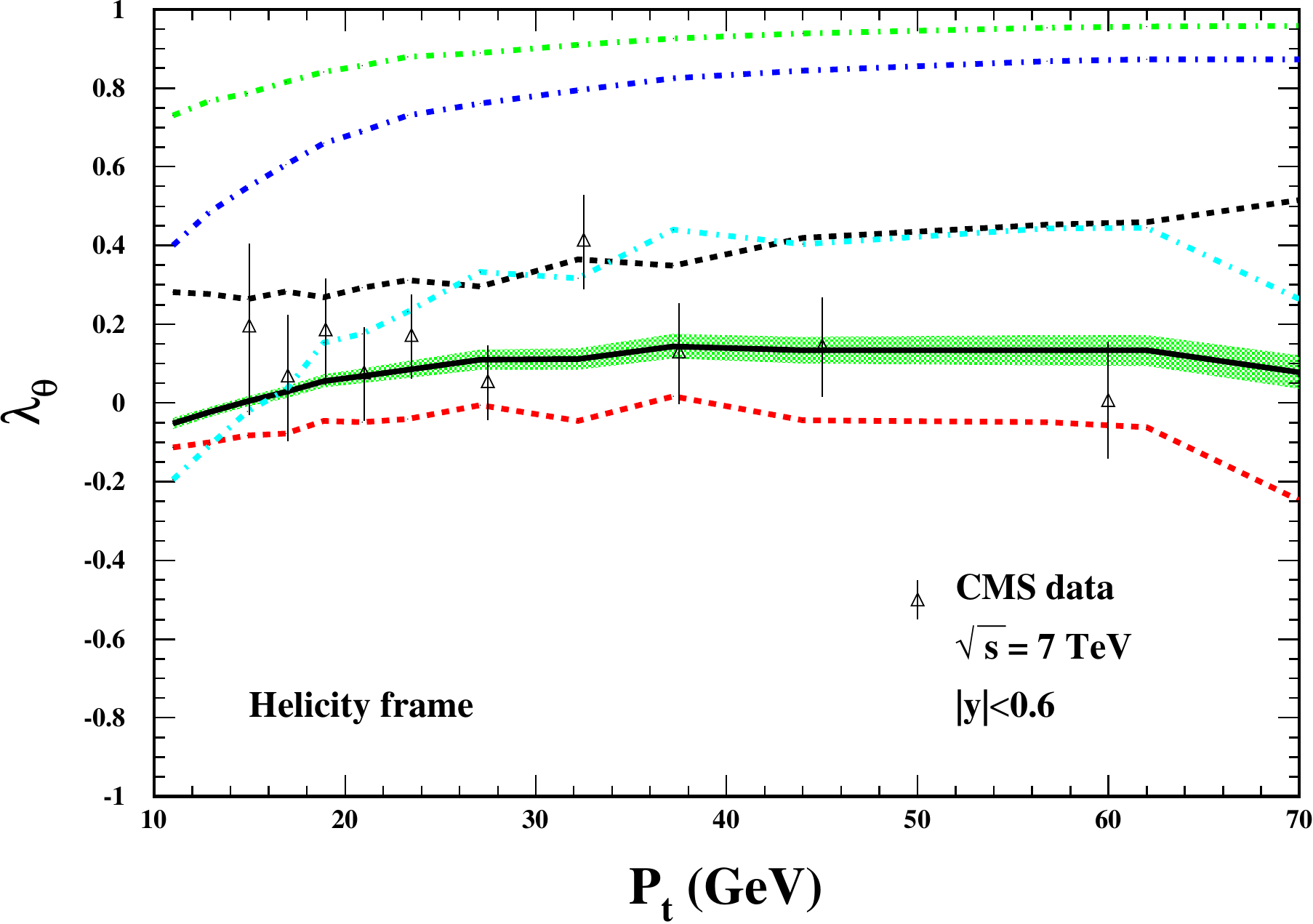}
  \includegraphics[width=0.3\textwidth,origin=c]{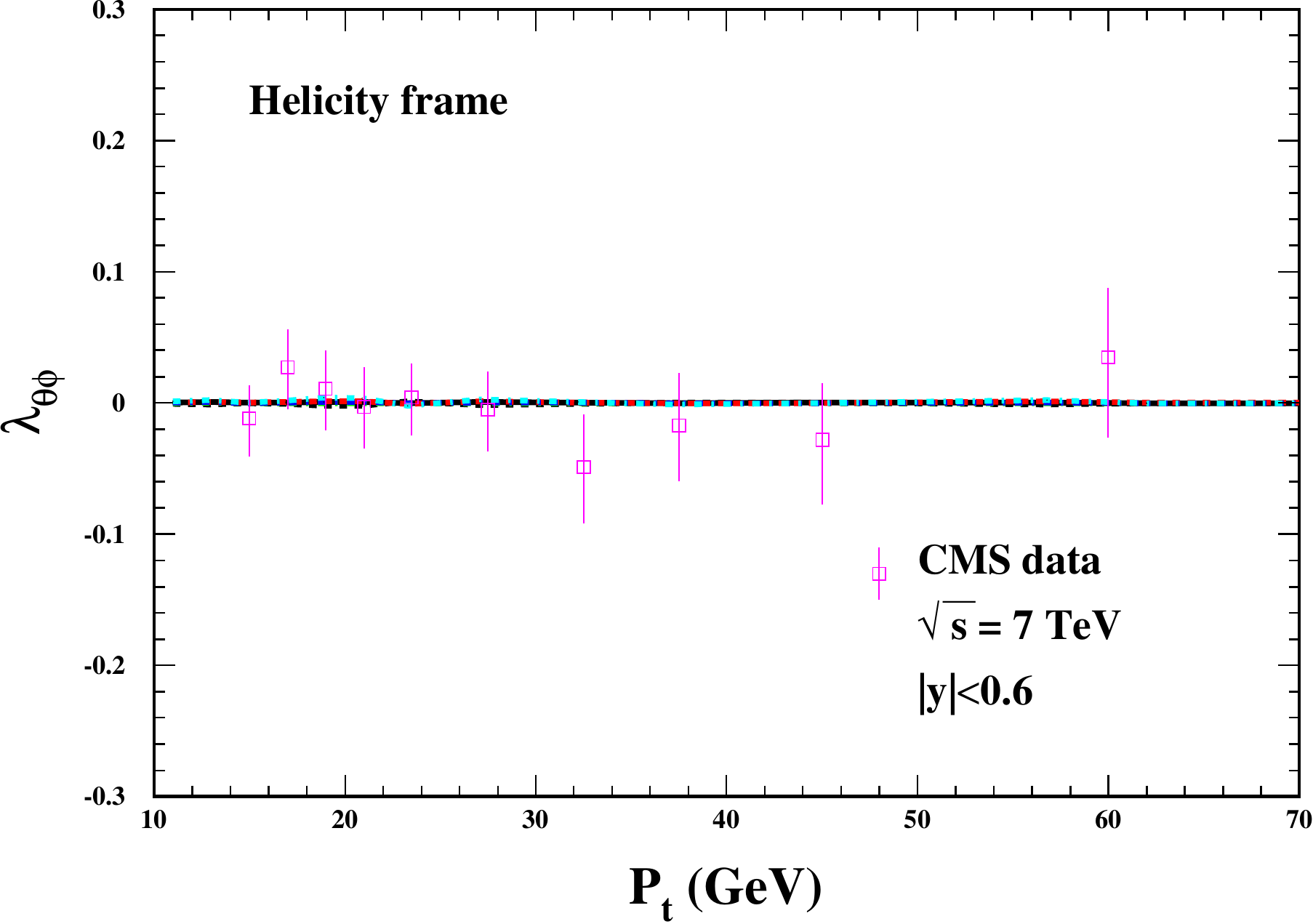}
  \includegraphics[width=0.3\textwidth,origin=c]{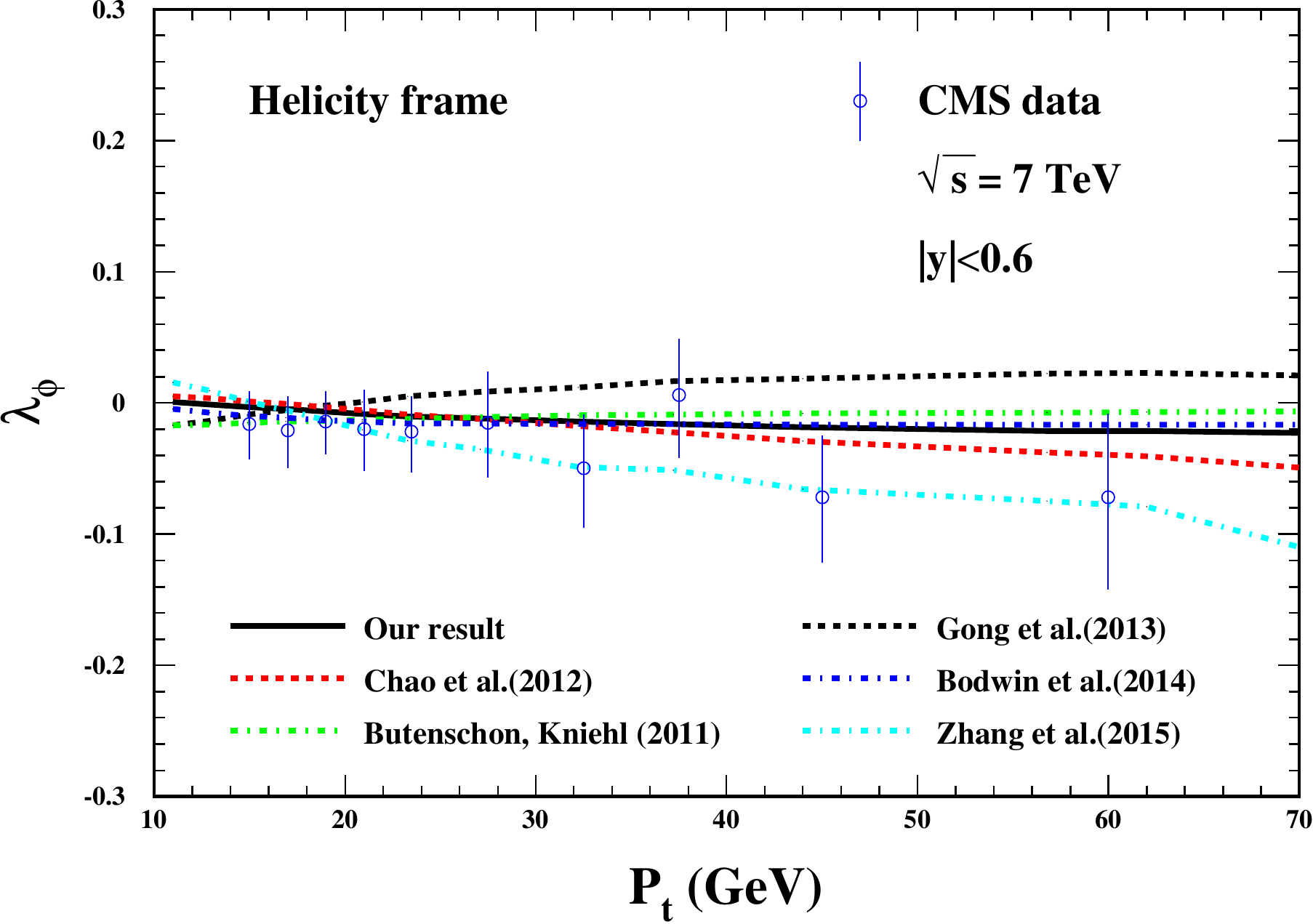}\\
  \includegraphics[width=0.3\textwidth,origin=c]{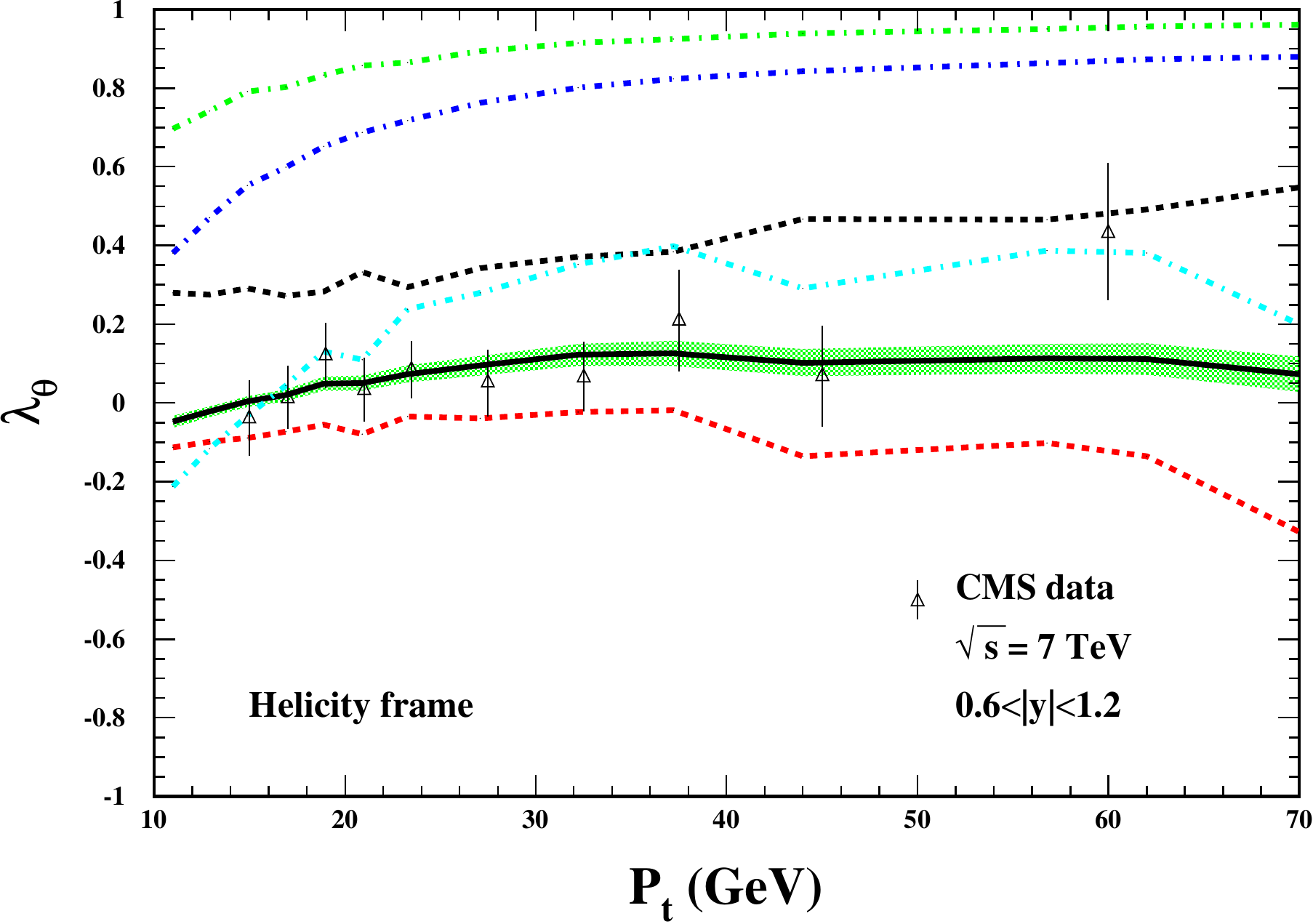}
  \includegraphics[width=0.3\textwidth,origin=c]{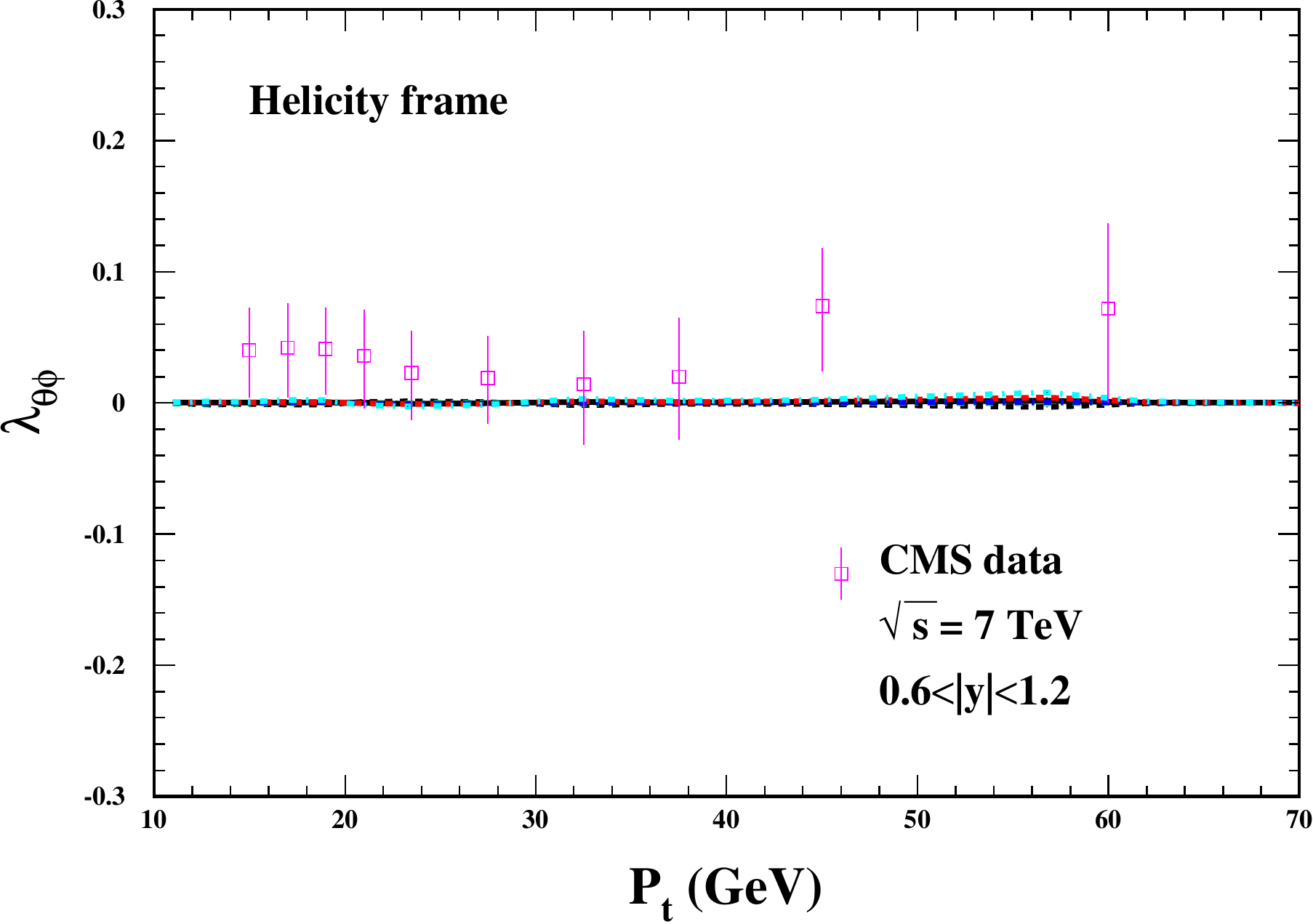}
  \includegraphics[width=0.3\textwidth,origin=c]{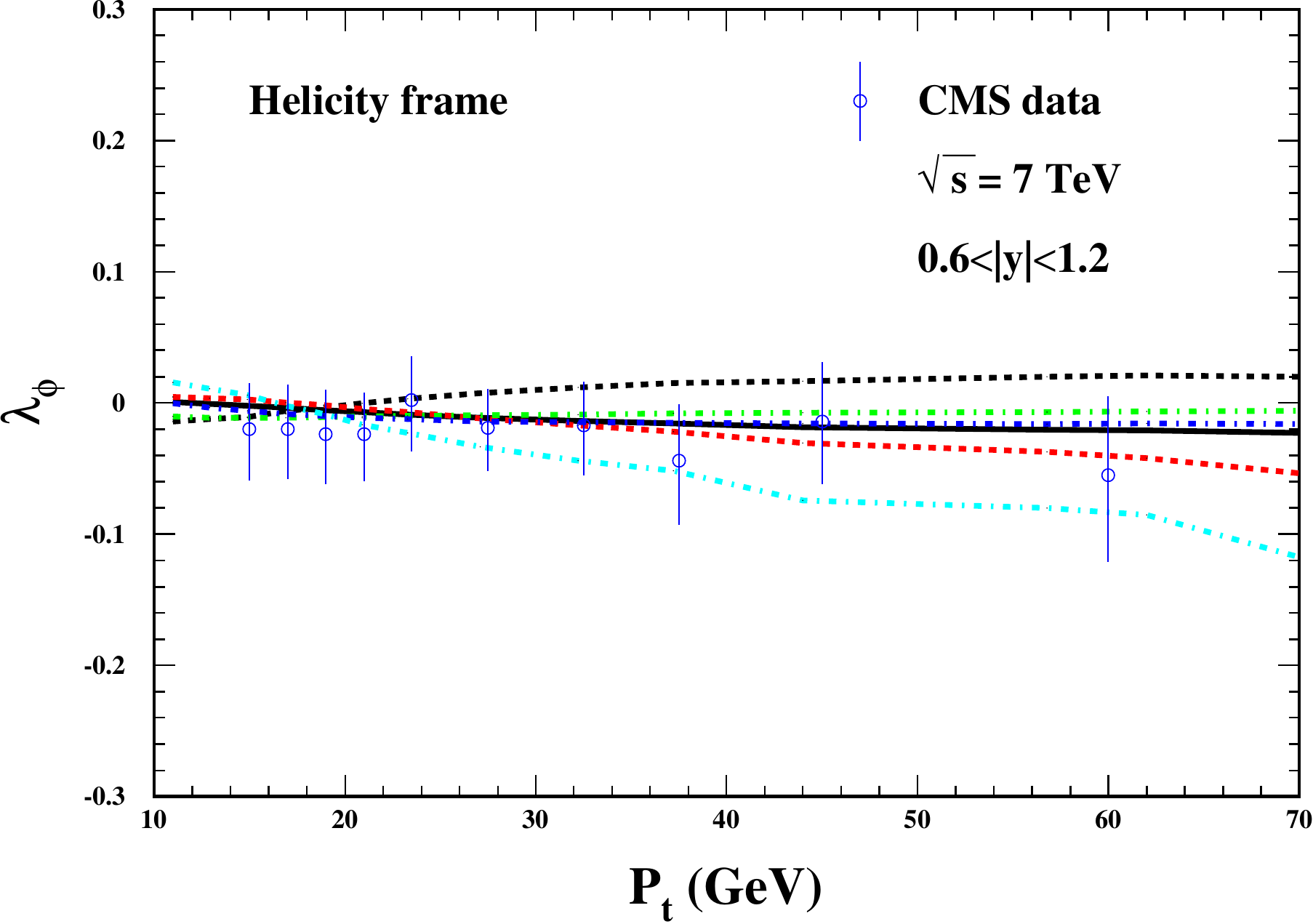}\\
\caption{\label{fig:cms-ldmes} $\lambda_{\theta}$ (left colum),
  $\lambda_{\theta\phi}$ (middle column) and $\lambda_{\phi}$ (right column)
  for $J/\psi$ with different sets of LDMEs
 at middle rapidity region. The CMS data is from Ref.~\cite{Chatrchyan:2013cla} }
\end{figure}

To investigate the uncertainties from different sets of LDMEs,
$\lambda_{\theta}$, $\lambda_{\theta\phi}$ and $\lambda_{\phi}$ are presented
by using the LDMEs sets in TABLE.~\ref{tab:ldmes-jpsi}.
In Fig.~\ref{fig:cms-ldmes},
the prediction of prompt $J/\psi$ polarization can explain the CMS data in a wide $p_t$ region
(10GeV $<p_t<$ 70GeV) in both $|y|<0.6$ and $0.6<|y|<1.2$ rapidity bins.
The newly fitted LDMEs provides an excellent description of the $\lambda_{\theta}$
at CMS window at transverse momentum range 14GeV $< p_t <$ 70GeV~\cite{Chatrchyan:2013cla}.
All the six fits scheme provide a good description of the $\lambda_{\theta\phi}$ and $\lambda_{\phi}$.
As we have mentioned in the above context, the SDC Re(d$\sigma_{10}$) is exact zero
in the symmetry rapidity region (e.g. $a\leq |y|\leq b$) for all the channels.
So the value of LDMEs would not alter the $\lambda_{\theta\phi}$ from zero for CMS data.
The CMS data of $\lambda_{\phi}$ are covered by the uncertainty from six LDMEs fit schemes.

\begin{figure}[tbp]
  \centering
  \includegraphics[width=0.3\textwidth,origin=c]{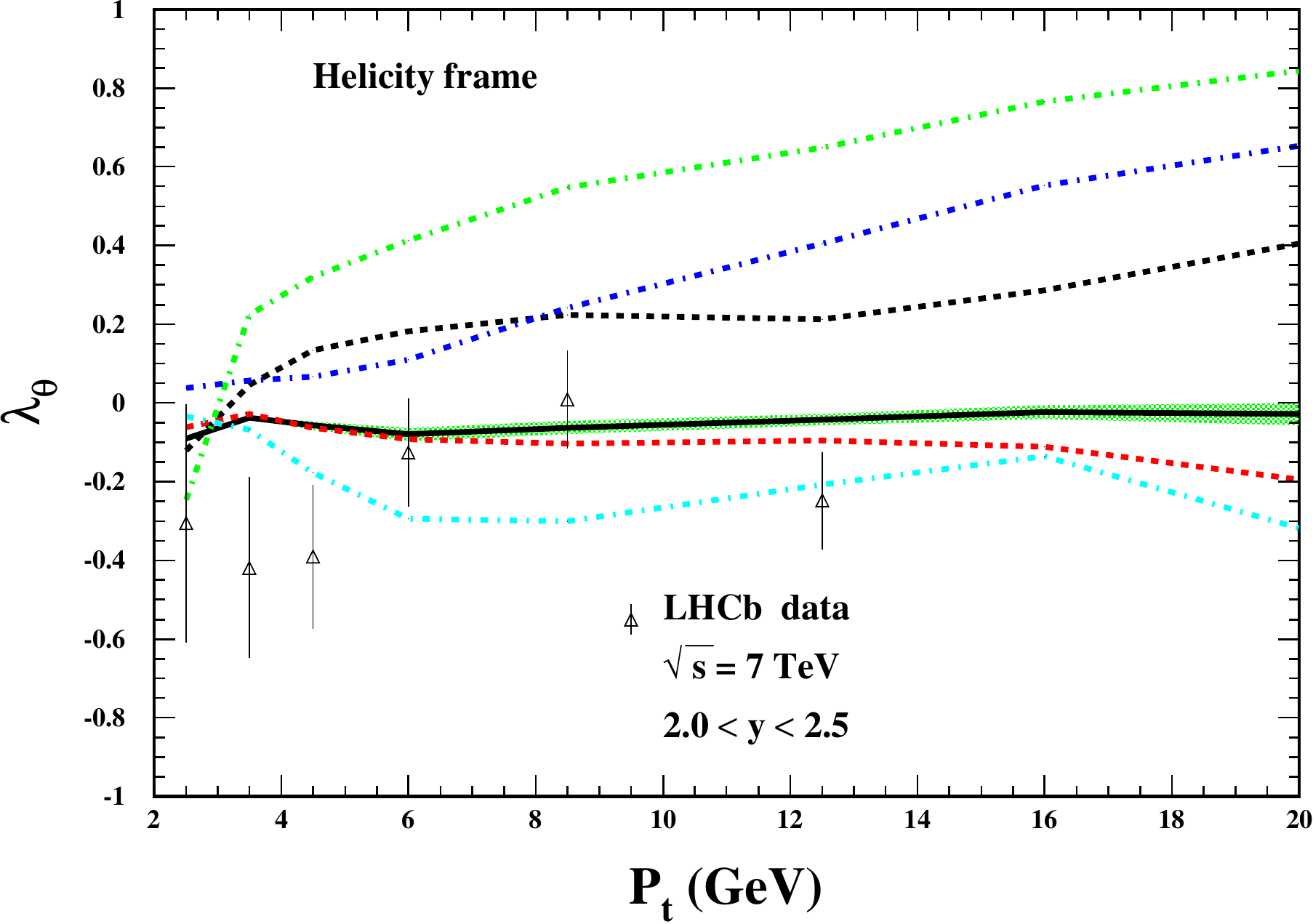}
  \includegraphics[width=0.3\textwidth,origin=c]{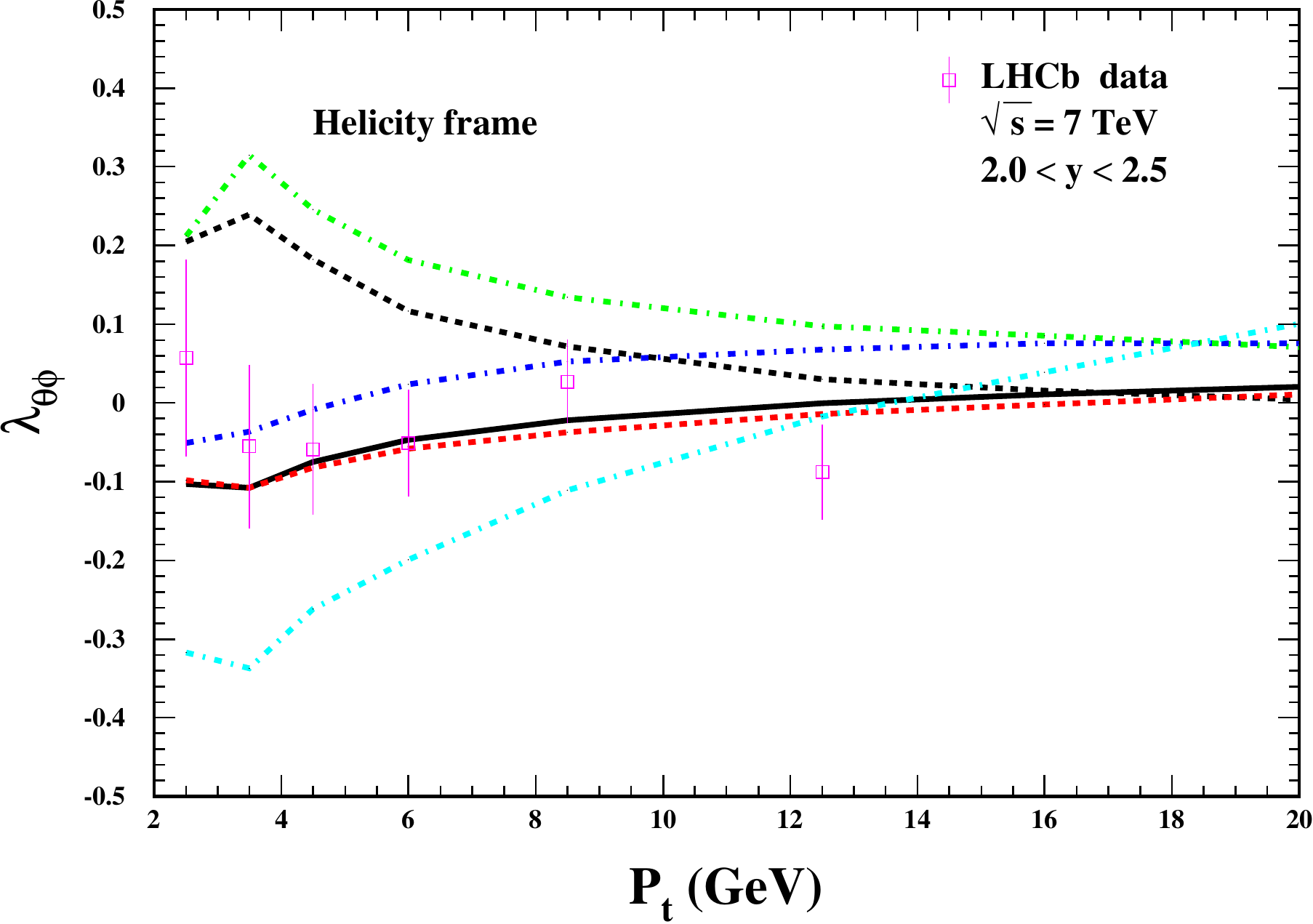}
  \includegraphics[width=0.3\textwidth,origin=c]{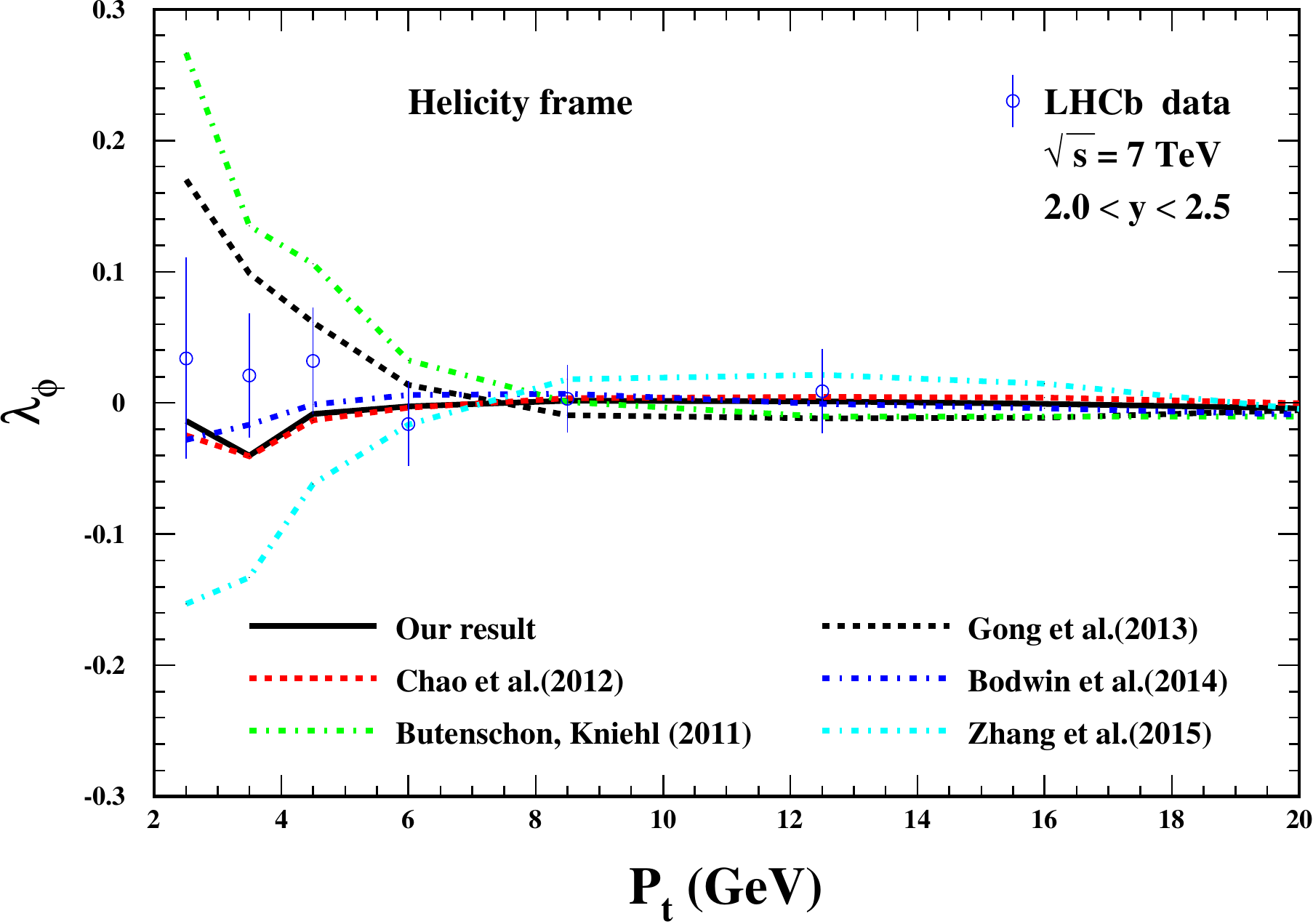}\\
  \includegraphics[width=0.3\textwidth,origin=c]{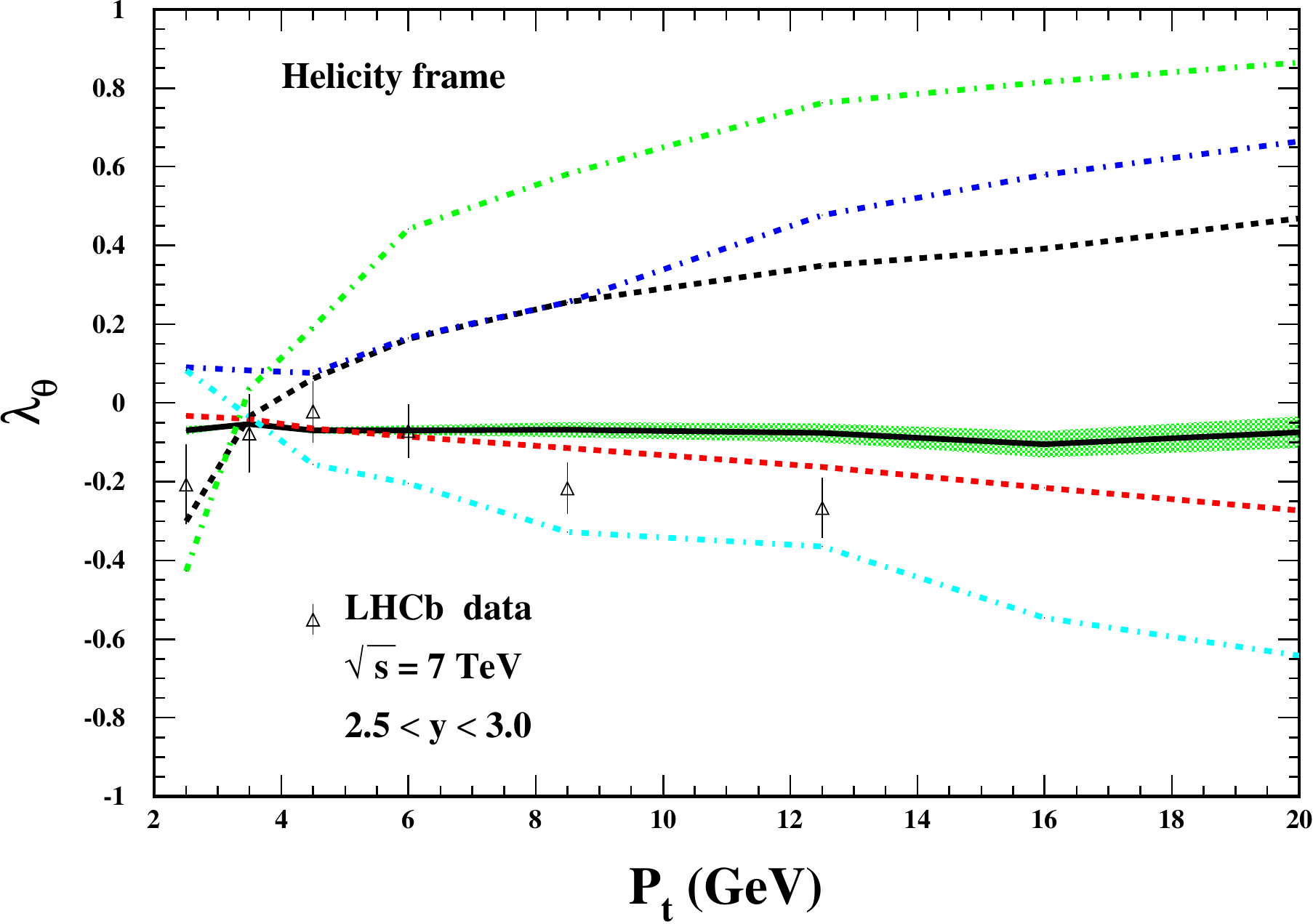}
  \includegraphics[width=0.3\textwidth,origin=c]{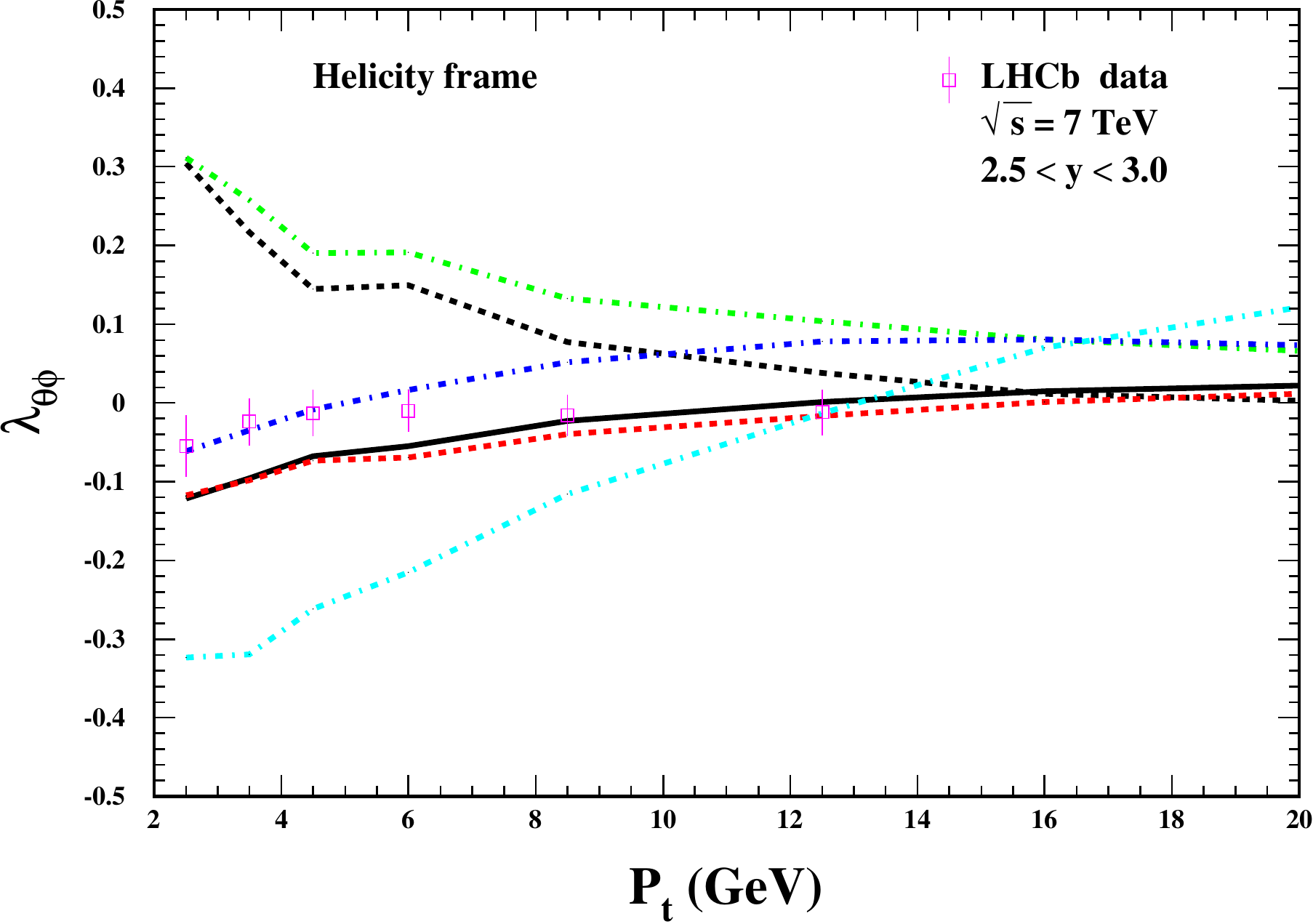}
  \includegraphics[width=0.3\textwidth,origin=c]{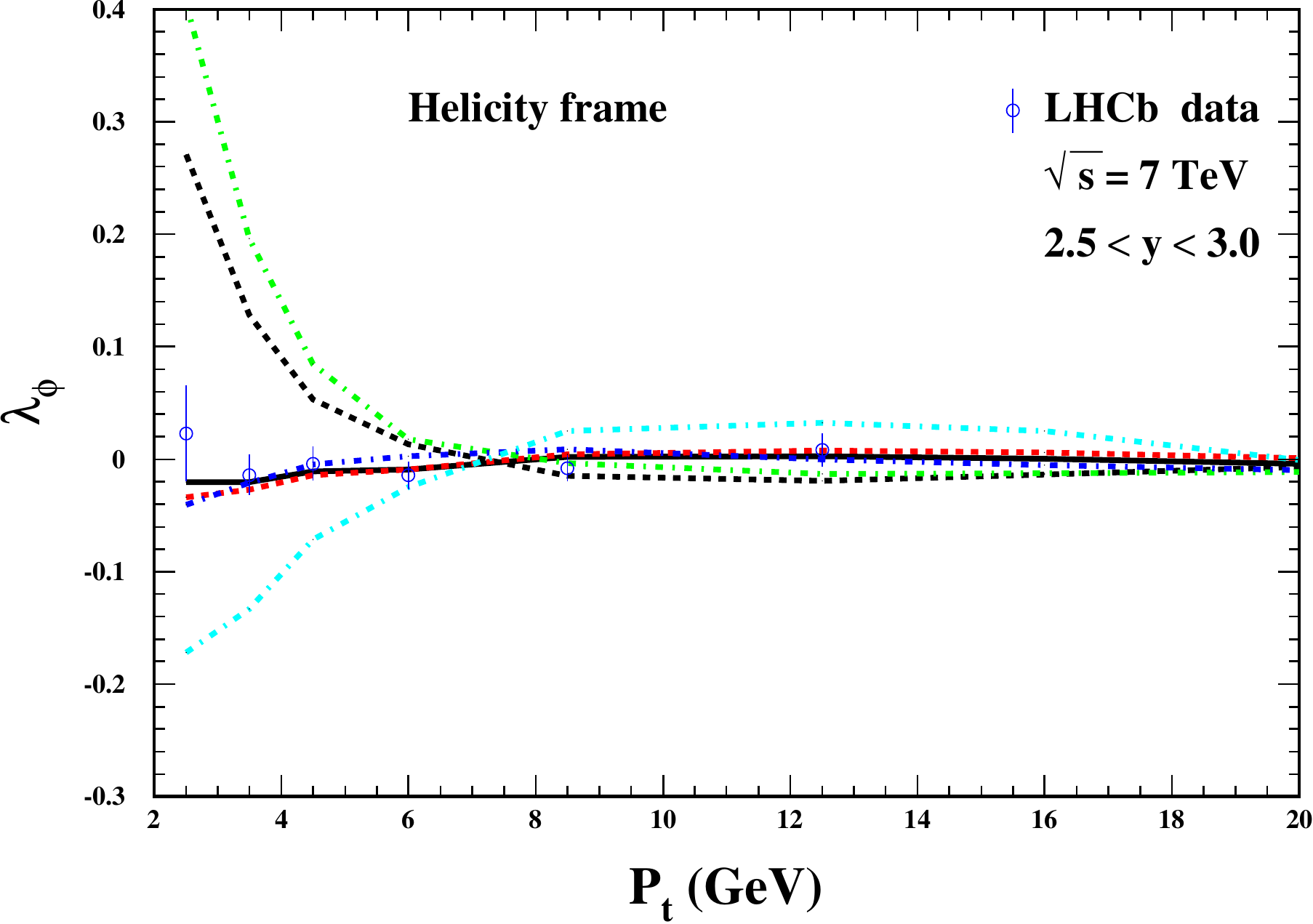}\\
  \includegraphics[width=0.3\textwidth,origin=c]{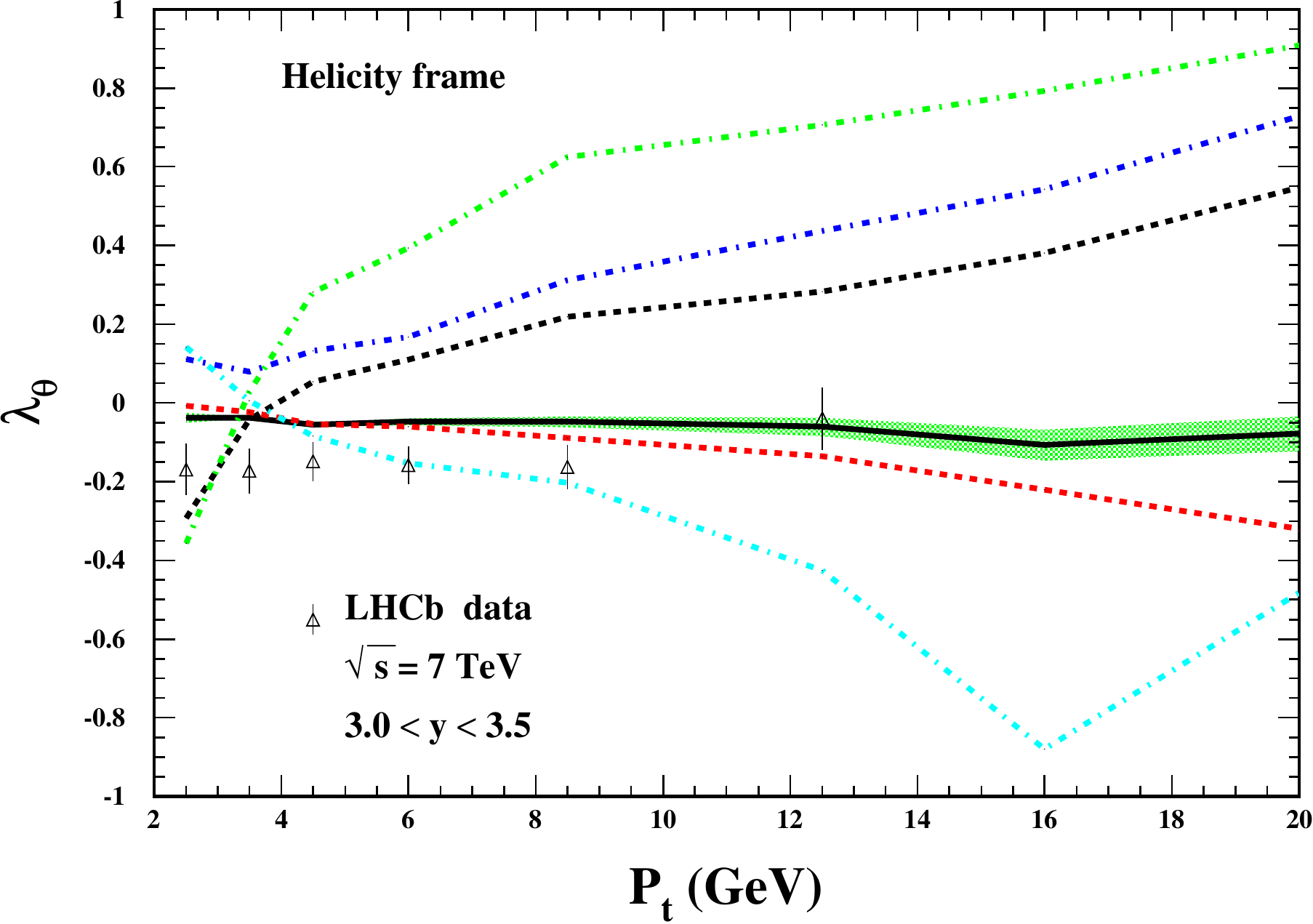}
  \includegraphics[width=0.3\textwidth,origin=c]{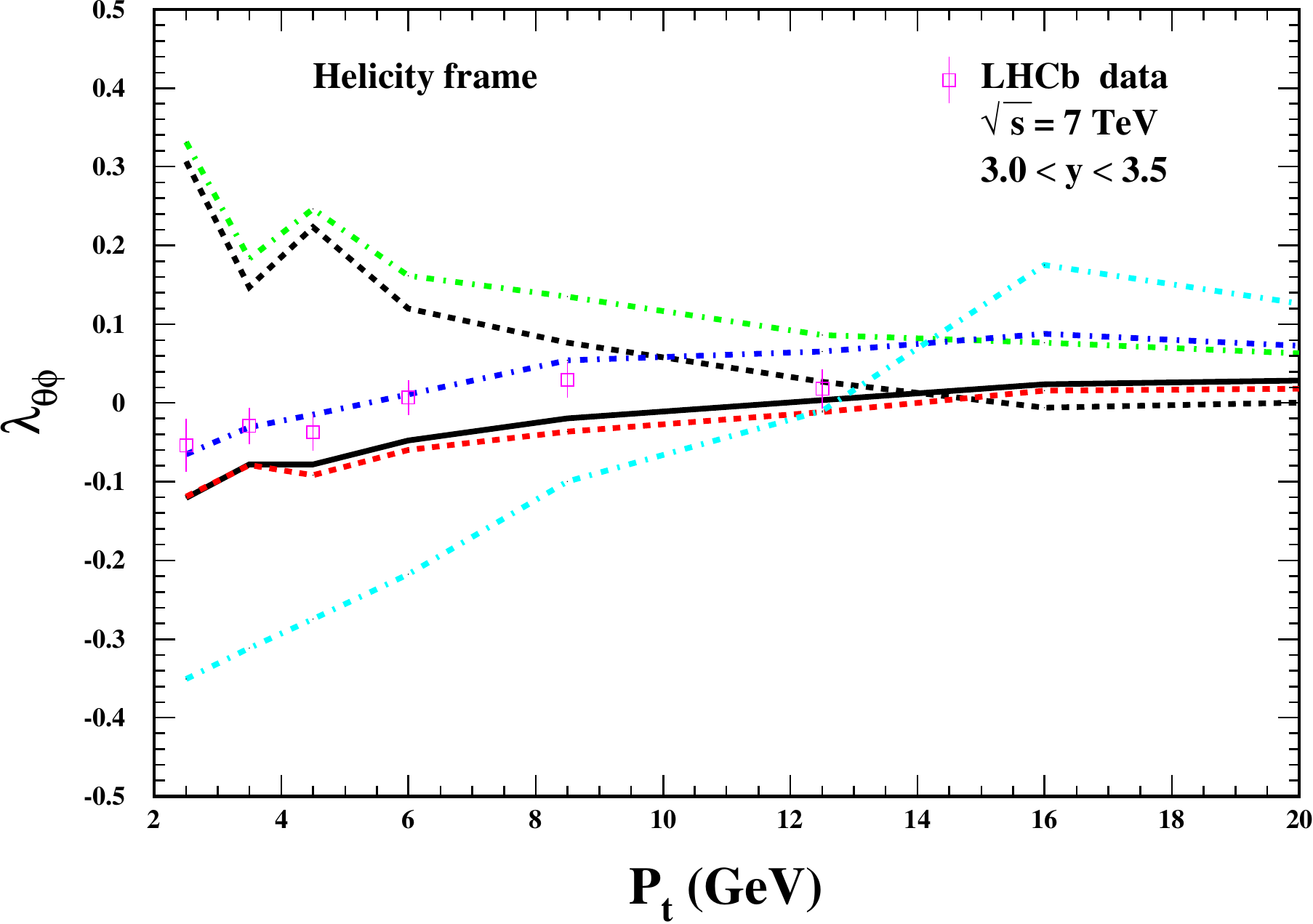}
  \includegraphics[width=0.3\textwidth,origin=c]{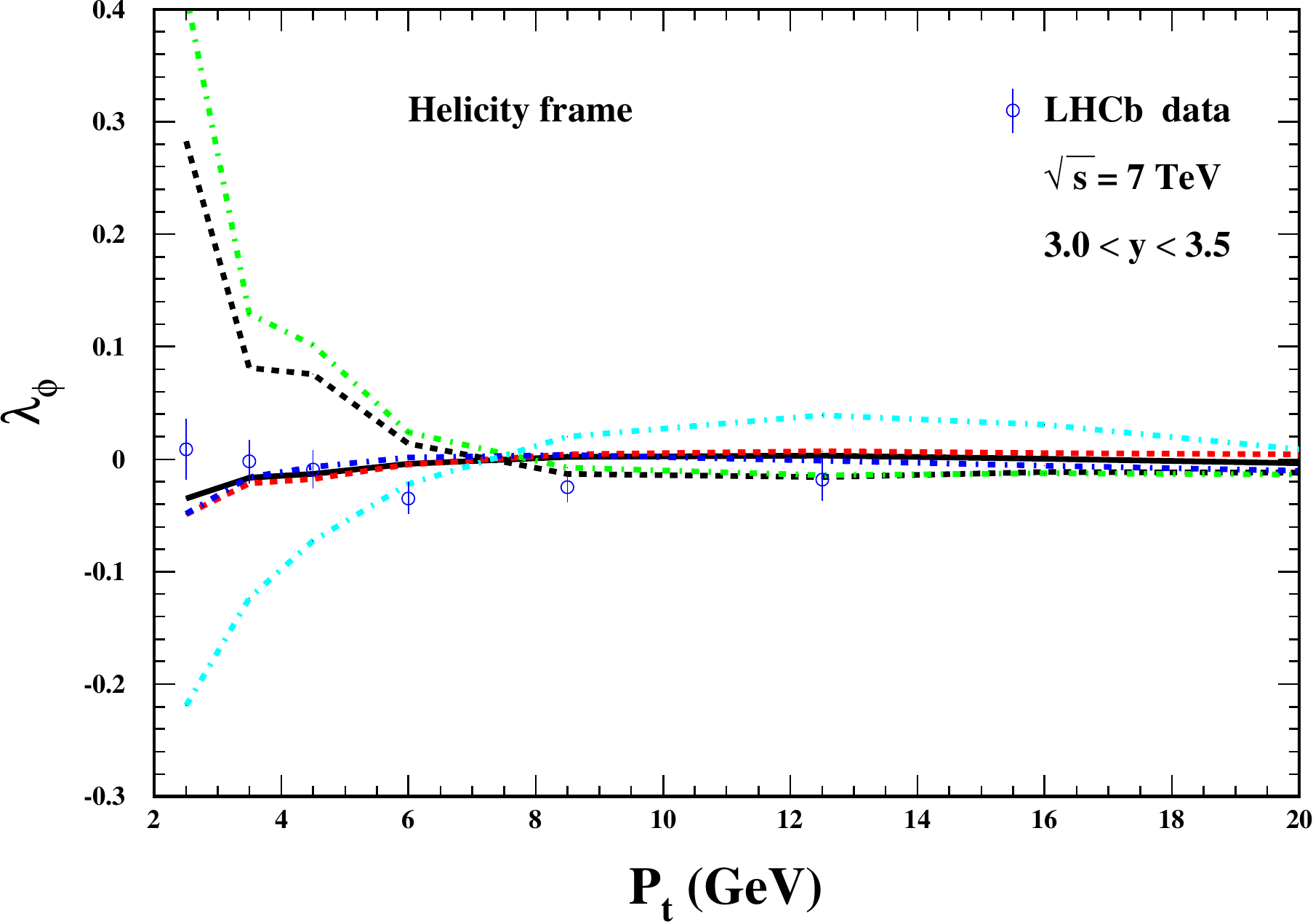}\\
  \includegraphics[width=0.3\textwidth,origin=c]{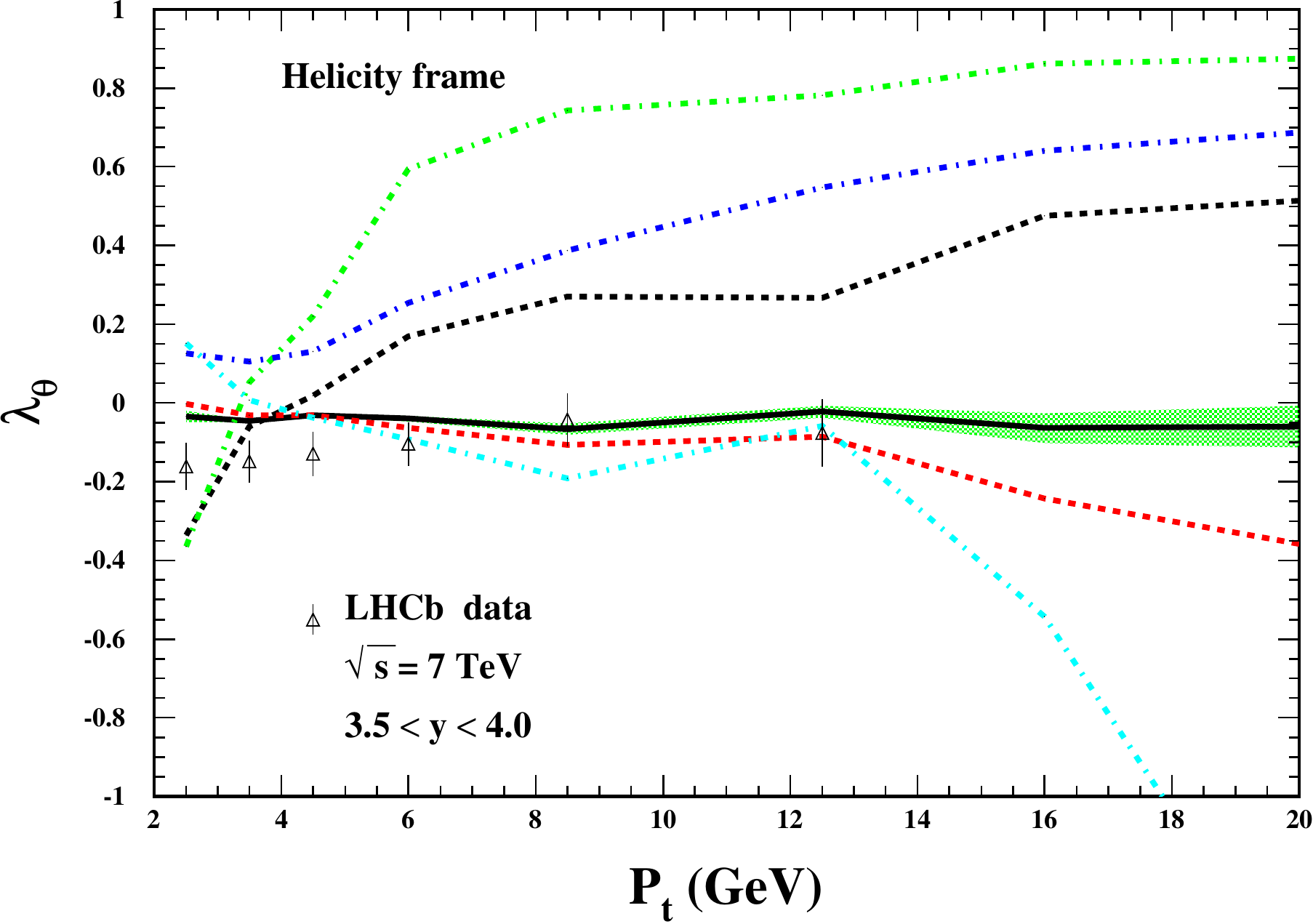}
  \includegraphics[width=0.3\textwidth,origin=c]{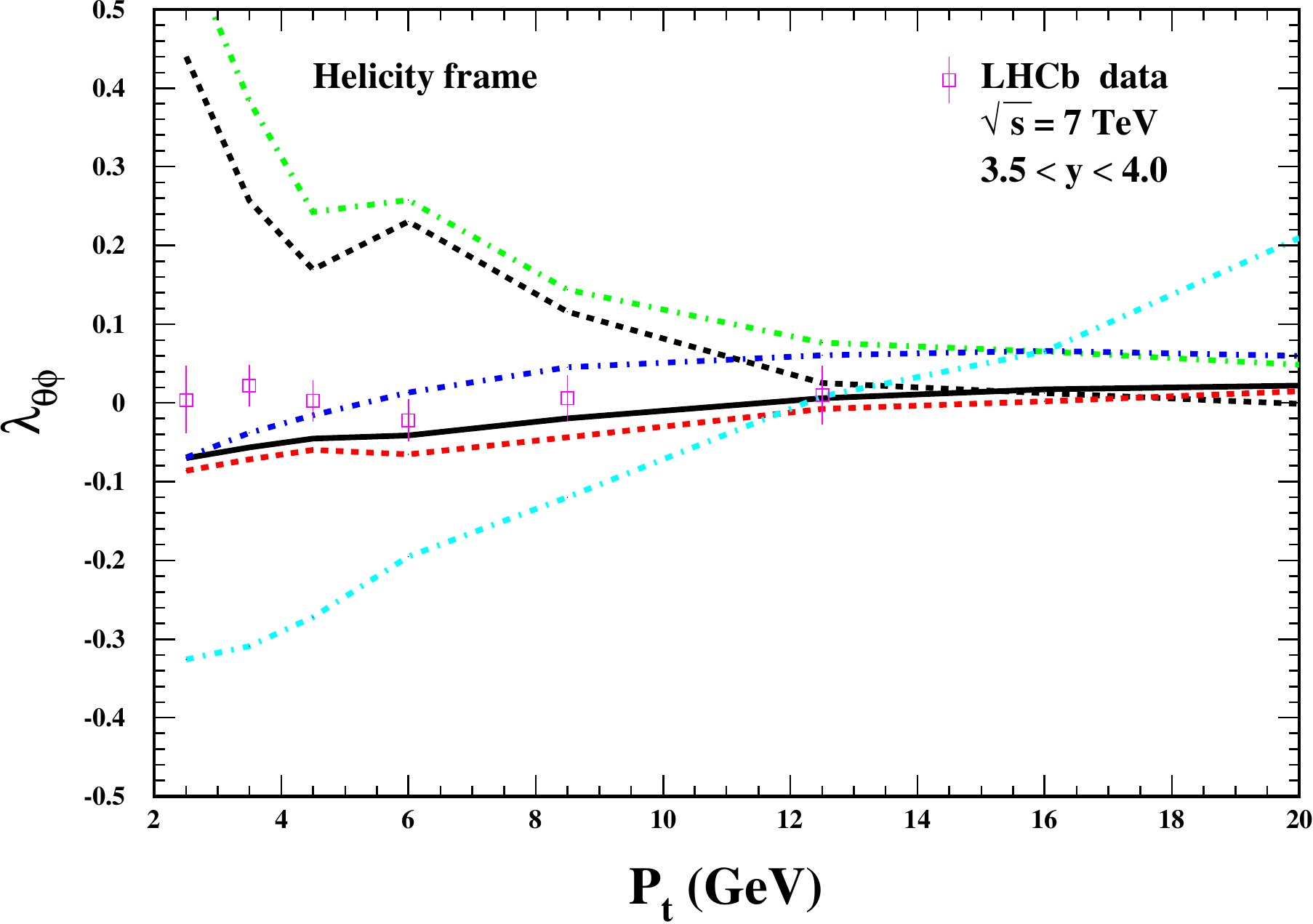}
  \includegraphics[width=0.3\textwidth,origin=c]{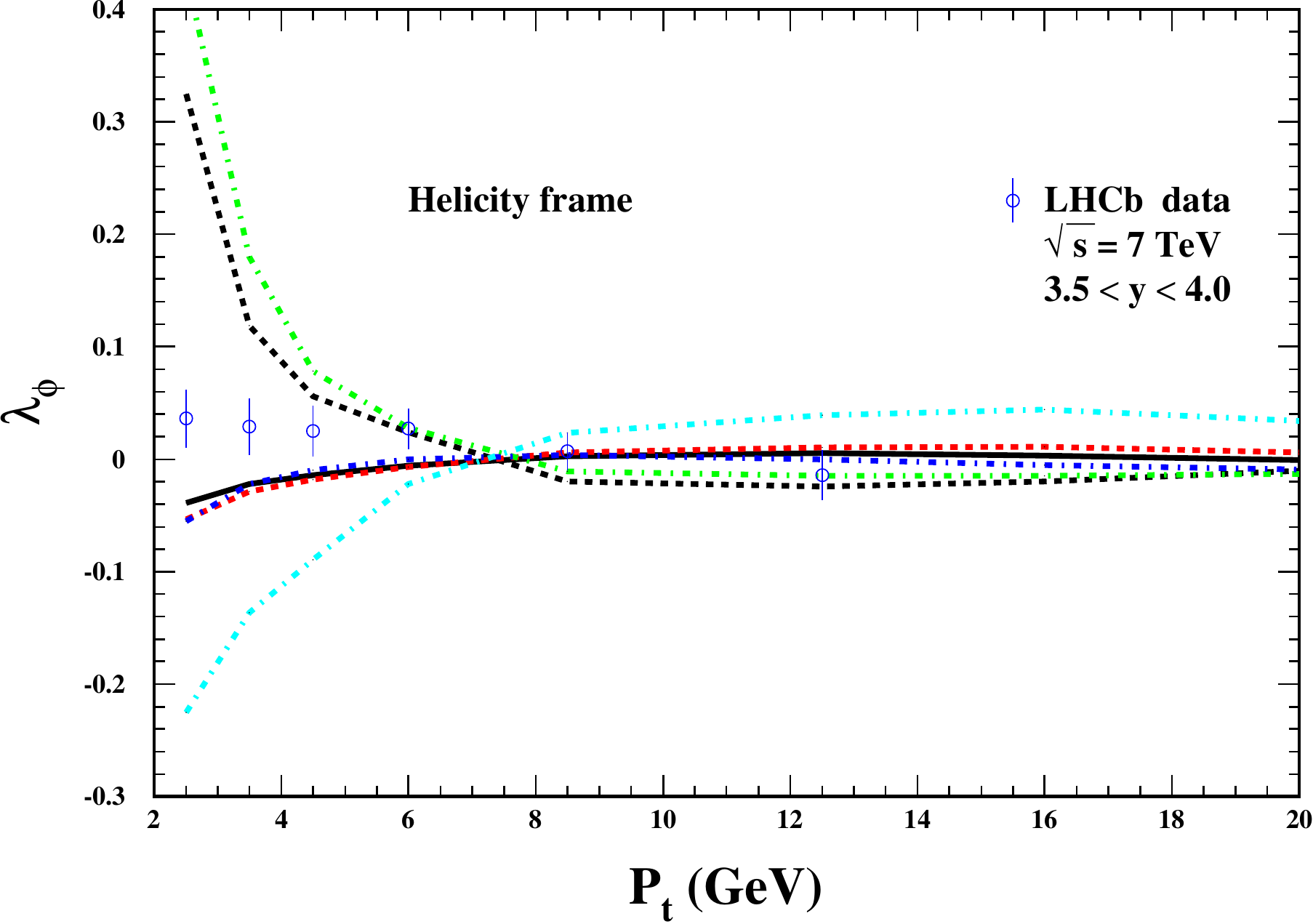}\\
  \includegraphics[width=0.3\textwidth,origin=l]{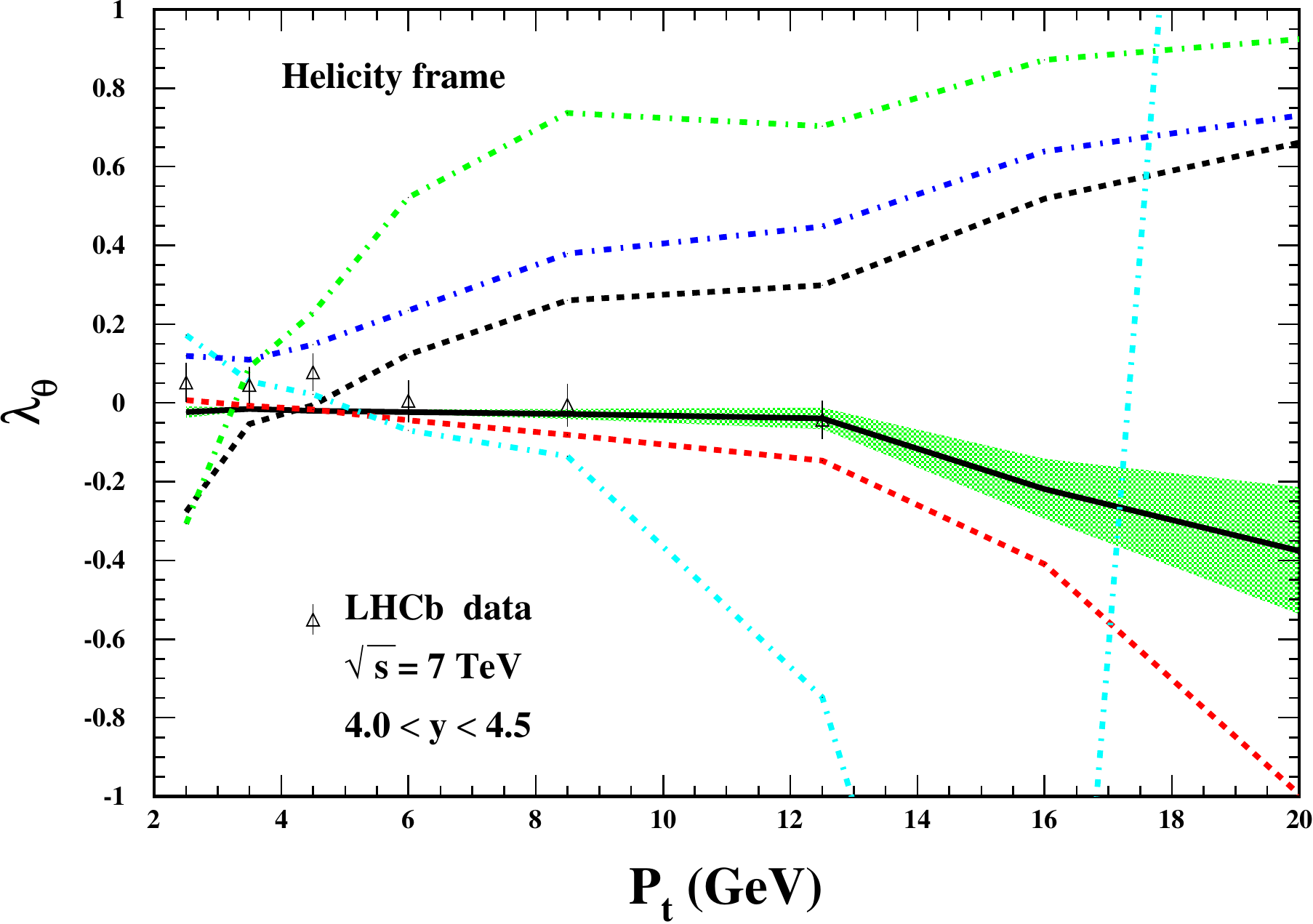}
  \includegraphics[width=0.3\textwidth,origin=l]{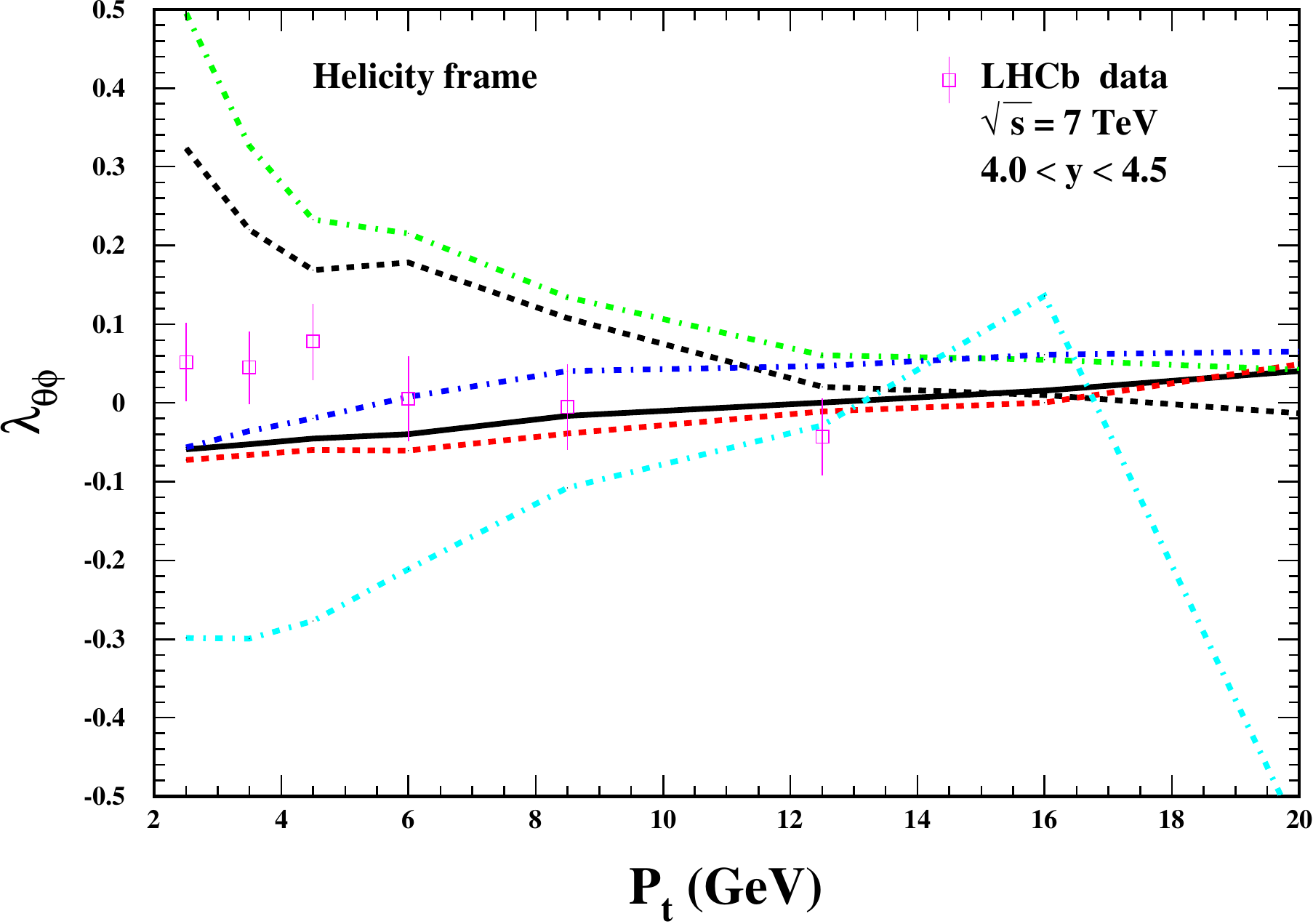}
  \includegraphics[width=0.3\textwidth,origin=l]{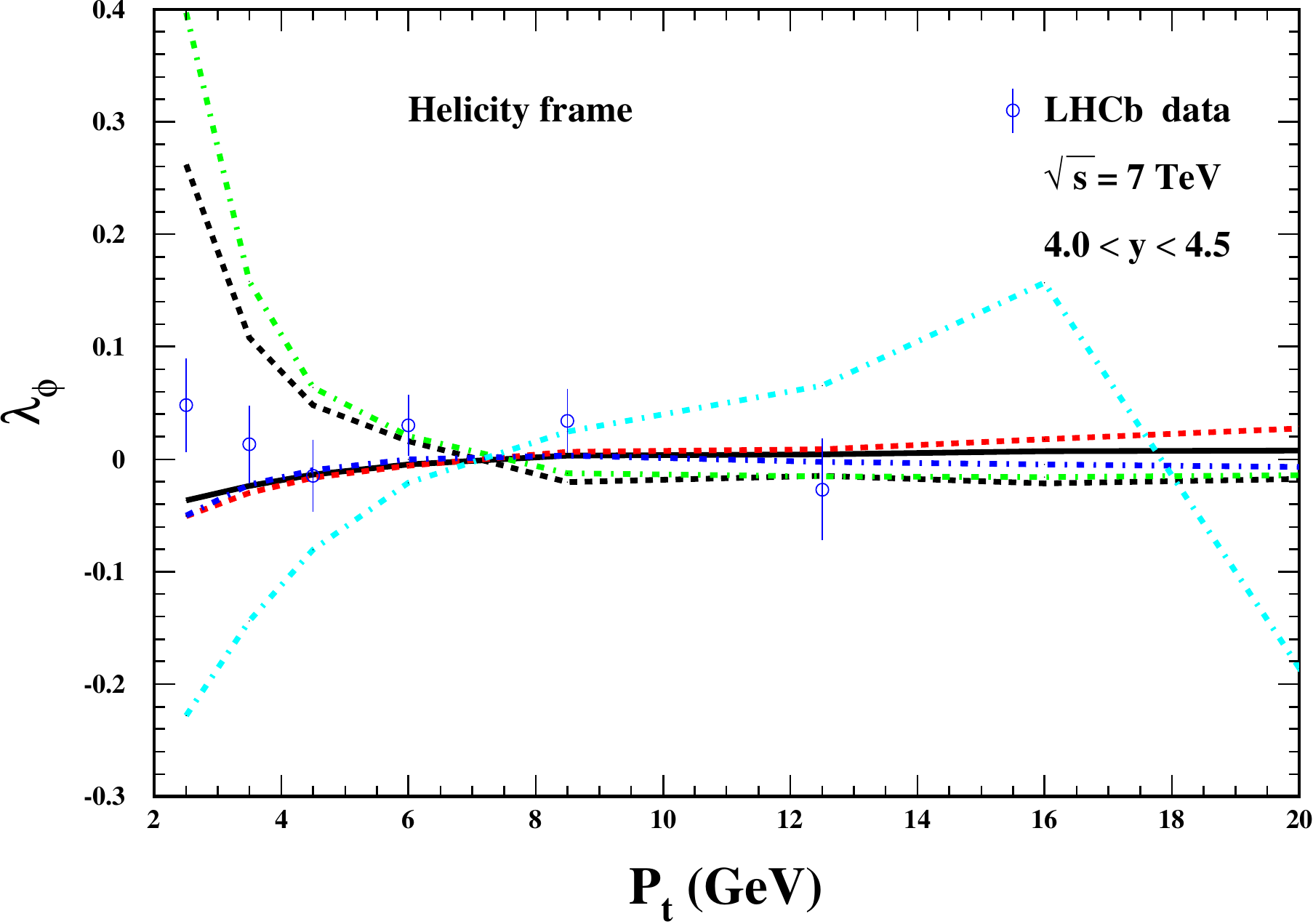}\\
\caption{\label{fig:lhc-ldmes} $\lambda_{\theta}$ (left colum),
  $\lambda_{\theta\phi}$ (middle column)
  and $\lambda_{\phi}$ (right column) for $J/\psi$ with
  different sets of LDMEs at forward rapidity region, where
  rows from top to bottom correspond to the rapidity ranges
  $2.<y<2.5$, $2.5<y<3.$, $3.<y<3.5$, $3.5<y<4.$, $4.<y<4.5$, respectively.
  The LHCb data is from Ref.~\cite{Aaij:2013nlm}.}
\end{figure}

In Fig.~\ref{fig:lhc-ldmes}, $\lambda_{\theta}$, $\lambda_{\theta\phi}$ and $\lambda_{\phi}$
in five rapidity bins at the LHCb are presented
from the top to the bottom rows.
For the different rapidity bins,
the $p_t$ distribution of $\lambda_{\theta}$, $\lambda_{\theta\phi}$ and $\lambda_{\phi}$ behave in a similar way.
At low $p_t$ region ($p_t < 4$ GeV), the different LDMEs sets provide large uncertainty,
while $p_t$ increasing, the results from the different LDMEs sets converge to zero.
It clearly shown that our new fit coincide with the experimental measurements at the LHCb quite well for $J/\psi$ production.

{\it \textbf{Summary and conclusion}---}
In this Letter, we finished
calculation on $\lambda_{\theta\phi}$ and $\lambda_{\phi}$
for $J/\psi$ and $\psi(2S)$ polarization in helicity frame based on NRQCD at QCD NLO.
The results are obtained for the first time and they coincide with the experimental measurements at the LHC quite well.
Therefore, the last two pieces for the $J/\psi$ polarization are successfully explained.
It means that the long-standing $J/\psi$ polarization puzzle is settle down completely.

By applying P parity invariance analysis, we obtained the conclusions in helicity frame
that $\lambda_{\theta\phi}=0$ for experiment
with symmetry rapidity range ($a<|y|<b$) like that at the CMS and ATLAS,
and $\lambda_{\theta\phi}\neq0$ for half rapidity range ($y>b$) such as the case at the LHCb
for polarization of $J/\psi$ and $\psi(2S)$ hadroproduction at the LHC.
However,  the electro-weak production for $J/\psi$ can break P-parity, our calculation shown that the leading contribution from
$^3S_1^{[1]}$, $^1S_0^{[8]}$, $^3S_1^{[8]}$, $^3P_J^{[8]}$ is smaller about 5 order in magnitude than the QCD
production processes. Therefore this contribution can be ignored.
This conclusion could be applied in the CMS or ATLAS experimental measurement to keep $\lambda_{\theta\phi}=0$
in each iteration of fitting to improve the measurements. It also indicate that the polarization
measurement could be performed in half rapidity range with $y>0$ or $y<0$ separately, in this way, $\lambda_{\theta\phi}\neq0$
could be achieved.

{\it \textbf{acknowledgments}---}
The results described here are achieved by using HPC Cluster of ITP-CAS. The author (Yu Feng) would like to thank CAS Key
Laboratory of Theoretical Physics, Institute of Theoretical Physics (ITP) CAS, for very kind invitation and part of this
work was completed during his visit of ITP. This work was supported in part by the National Natural Science Foundation of
China with Grants Nos. 11747037, 11275243, 11275036, 11447601, 11535002, 11675239 and 11475183. It is also supported by Key Research
Program of Frontier Sciences, CAS, Grant No. QYZDY-SSW-SYS006. Y. Feng was also supported by the Army Medical University
of PLA of China No.2016XPY06.

\bibliography{paper.bbl}

\end{document}